\pdfoutput=1

\documentclass[11pt,twoside,a4paper,cmspaper,final,collab]{cms-tdr}
\def\svnVersion{479861P}\def\cmsCernNoTag{CERN-EP-2018-228}\def\cmsCernDate{\today}\def\cmsMessage{Published in the Journal of High Energy Physics as \href{http://dx.doi.org/10.1007/JHEP10(2018)138}{\doi{10.1007/JHEP10(2018)138.}}}

\begin{document}\cmsNoteHeader{HIN-18-004}

\hyphenation{had-ron-i-za-tion}
\hyphenation{cal-or-i-me-ter}
\hyphenation{de-vices}
\RCS$HeadURL: svn+ssh://svn.cern.ch/reps/tdr2/papers/HIN-18-004/trunk/HIN-18-004.tex $
\RCS$Id: HIN-18-004.tex 477895 2018-10-12 12:26:06Z abaty $
\newlength\cmsTabSkip
\setlength\cmsTabSkip{1.5ex}
\providecommand{\NA}{\ensuremath{\text{---}}}
\providecommand{\CL}{CL\xspace}

\newcommand{\sqrts}{\ensuremath{\sqrt{s}}\xspace}
\newcommand{\PbPb}{\ensuremath{\text{PbPb}}\xspace}
\newcommand{\XeXe}{\ensuremath{\text{XeXe}}\xspace}
\newcommand{\raa}{\ensuremath{R_{\text{AA}}}\xspace}
\newcommand{\raaStar}{\ensuremath{R^{*}_{\text{AA}}}\xspace}
\newcommand{\rxepb}{\ensuremath{R_{\text{Pb}}^{\text{Xe}}}\xspace}
\newcommand{\avgNColl}{$\langle N_{\text{coll}} \rangle$\xspace}
\newcommand{\avgNPart}{$\langle N_{\text{part}} \rangle$\xspace}
\newcommand{\TAA}{\ensuremath{T_{\text{AA}}}\xspace}

\cmsNoteHeader{HIN-18-004}
\title{Charged-particle nuclear modification factors in \XeXe collisions at $\sqrtsNN=5.44\TeV$}
\date{\today}

\abstract{
The differential yields of charged particles having pseudorapidity within $\abs{\eta}<1$ are measured using xenon-xenon (\XeXe) collisions at $\sqrtsNN=5.44\TeV$. The data, corresponding to an integrated luminosity of 3.42\mubinv, were collected in 2017 by the CMS experiment at the LHC.  The yields are reported as functions of collision centrality and transverse momentum, \pt, from 0.5 to 100\GeV.  A previously reported \pt spectrum from proton-proton collisions at $\sqrt{s}=5.02\TeV$ is used for comparison after correcting for the difference in center-of-mass energy.  The nuclear modification factors using this reference, \raaStar, are constructed and compared to previous measurements and theoretical predictions.  In head-on collisions, the \raaStar has a value of 0.17 in the \pt range of 6--8\GeV, but increases to approximately 0.7 at 100\GeV. Above ${\approx}6\GeV$, the \XeXe data show a notably smaller suppression than previous results for lead-lead (\PbPb) collisions at $\sqrtsNN=5.02\TeV$ when compared at the same centrality (\ie, the same fraction of total cross section). However, the \XeXe suppression is slightly greater than that for \PbPb in events having a similar number of participating nucleons.
}

\hypersetup{%
pdfauthor={CMS Collaboration},%
pdftitle={Charged-particle nuclear modification factors in XeXe collisions at sqrt(sNN)=5.44 TeV},%
pdfsubject={CMS},%
pdfkeywords={CMS, physics, heavy ions }}

\maketitle \section{Introduction}
\label{sec:intro}
The transverse momentum (\pt) spectrum of charged particles is a well-studied observable for examining the hot, dense quark-gluon plasma (QGP) created in high-energy heavy ion collisions.  As scattered partons traverse this medium, they experience a loss of energy due to quantum chromodynamics processes such as gluon emission and parton splitting~\cite{Qin:2015srf}.  Because high-\pt charged particles are produced through parton fragmentation and subsequent hadronization, their yields are sensitive to the strength of QGP-induced energy loss~\cite{Bjorken:1982tu,d'Enterria:2009am}. In contrast, production of charged particles having \pt less than a few \GeV is particularly sensitive to initial parton densities and hydrodynamic expansion of the medium~\cite{Gyulassy:2000gk,Poskanzer:1998yz,Ollitrault:1992bk,Kolb:2000sd}.

Modification of charged-particle yields can be quantified by forming a ratio of the spectra in nucleus-nucleus (AA) and $\Pp\Pp$ collisions, where the latter are multiplied by the average number of binary nucleon-nucleon collisions per AA event, \avgNColl.  This observable is known as the nuclear modification factor, \raa, and is given by
\begin{linenomath}
\begin{equation}
R_{\text{AA}}(\pt) = \frac{1}{\langle N_{\text{coll}}\rangle }\frac{\rd N^{\text{AA}}/\rd \pt}{\rd {N^{\Pp\Pp}/\rd \pt}}.
\label{eqn:def_raa}
\end{equation}
\end{linenomath}
Here $\rd N^{\text{AA}}/\rd \pt$ ($\rd {N^{\Pp\Pp}}/\rd \pt$) is the charged-particle yield in AA ($\Pp\Pp$) collisions.  An equivalent definition replaces $\rd {N^{\Pp\Pp}}/\rd \pt$ with the differential charged-particle cross section in inelastic $\Pp\Pp$ collisions, $\rd {\sigma^{\Pp\Pp}}/\rd \pt$, and \avgNColl with the nuclear overlap function, $\TAA=\langle N_{\text{coll}} \rangle/\sigma^{\Pp\Pp}$:
\begin{linenomath}
\begin{equation}
\label{eqn:def_raa2}
R_{\text{AA}}(\pt) = \frac{1}{\TAA}\frac{\rd N^{\text{AA}}/\rd \pt}{\rd \sigma^{\Pp\Pp}/\rd \pt}.
\end{equation}
\end{linenomath}
Both \avgNColl and \TAA can be obtained using a Glauber model of nuclear collisions~\cite{Miller:2007ri}.

Charged-particle \pt spectra and their associated nuclear modification have been explored at the BNL RHIC~\cite{Arsene:2004fa,Back:2004je,Adams:2005dq,Adcox:2004mh} in gold-gold collisions at a center-of-mass energy per nucleon pair ($\sqrtsNN$) of up to 200~\GeV.  These analyses found \raa to be strongly suppressed in head-on collisions, with minima around $\pt=5\GeV$.  Measurements made at the CERN LHC by the ALICE~\cite{Abelev:2012hxa,Acharya:2018qsh}, ATLAS~\cite{Aad:2015wga}, and CMS~\cite{CMS:2012aa,Khachatryan:2016odn} Collaborations have explored the same observables in lead-lead (\PbPb) collisions at $\sqrtsNN = 2.76$ and 5.02\TeV.  These studies found minima of \raa around 0.15 at $\pt=8\GeV$.  They also indicate that \raa increases to values around 0.7 at $\pt=100\GeV$.  Complementary measurements of the nuclear modification factor in proton-lead ($\Pp$Pb) collisions at $\sqrtsNN = 5.02\TeV$ indicate that high-\pt charged-particle yields are not strongly modified in this smaller colliding system, ruling out effects related to the initial-state conditions of the lead nucleus as a cause of the high-\pt suppression seen in \PbPb collisions~\cite{Acharya:2018qsh,Aad:2016zif,Khachatryan:2016odn}. Together, these observations indicate strong \pt-dependent energy loss due to the presence of the QGP in heavy ion collisions.

In 2017, the LHC collided $^{129}$Xe nuclei at $\sqrtsNN=5.44\TeV$. The LHC had previously only provided proton-proton ($\Pp\Pp$), $\Pp$Pb, and $\PbPb$ collisions.  Therefore, the xenon-xenon (\XeXe) data provide a unique opportunity to explore the properties of the QGP using an intermediate size collision system at LHC energies.  Xenon collisions also provide an opportunity to test the system size dependence of parton energy loss.  The radii of xenon and lead nuclei are ${\approx}5.4$ and ${\approx}6.6\unit{fm}$, respectively~\cite{Loizides:2014vua}.  Assuming the energy loss of a parton is linearly (quadratically) related to only its path length through the QGP would imply an average reduction in energy loss of 17 (31)\% in head-on \XeXe collisions as compared to \PbPb collisions. This difference could manifest itself in comparisons of the charged-particle spectra between the two systems.  Recent results from the ALICE Collaboration indicate this is the case, with the \raa of head-on XeXe collisions being less suppressed than that of PbPb collisions~\cite{Acharya:2018eaq}. Comparisons of copper-copper and gold-gold collisions at RHIC have also motivated similar conclusions~\cite{Arsene:2016myy, Alver:2005nb, Adare:2008ad, Abelev:2009ab}.

To facilitate comparison of these two collision systems, a scaled ratio between the \XeXe and \PbPb charged-particle spectra is defined as
\begin{linenomath}
\begin{equation}
R^{\text{Xe}}_{\text{Pb}}(\pt) = \frac{\rd N^{\text{XeXe}}/\rd \pt}{\rd N^{\text{PbPb}}/\rd \pt}\frac{T_{\text{PbPb}}}{T_{\text{XeXe}}}.
\label{eqn:RXePb}
\end{equation}
\end{linenomath}
Here the AA notation is replaced with the names of the appropriate ion species.  Unlike \raa, this ratio does not depend on $\Pp\Pp$ reference data. Because the \PbPb data were gathered at $\sqrtsNN=5.02\TeV$, the two collision systems compared in this paper have different center-of-mass energies.  A deviation of \rxepb from expected values, after taking this energy difference into account, would indicate a different spectral modification between \XeXe and \PbPb collisions.

In this paper, \pt spectra are reported for charged particles with pseudorapidity $\abs{\eta}<1$ in \XeXe collisions at $\sqrtsNN=5.44\TeV$.  A $\Pp\Pp$ reference spectrum at a center-of-mass energy ($\sqrts$) of 5.44\TeV is constructed by extrapolating from an existing measurement at $\sqrts = 5.02\TeV$~\cite{Khachatryan:2016odn}. This reference is used to estimate the nuclear modification factor \raaStar, where the asterisk denotes the use of an extrapolated reference.  The results for \raaStar are compared to theoretical calculations, and potential implications are discussed.

\section{The CMS detector}
\label{sec:detector}
The central feature of the CMS apparatus is a superconducting solenoid of 6\unit{m} internal diameter, providing a magnetic field of 3.8\unit{T}. Within the solenoid volume are a silicon pixel and strip tracker, a lead tungstate crystal electromagnetic calorimeter, and a brass and scintillator hadron calorimeter, each composed of a barrel and two endcap sections. Forward calorimeters extend the $\eta$ coverage provided by the barrel and endcap detectors. Muons are detected in gas-ionization chambers embedded in the steel flux-return yoke outside the solenoid.

The silicon tracker measures charged particles within the range $\abs{\eta} < 2.5$. It consists of 1856 silicon pixel and 15\,148 silicon strip detector modules. For nonisolated particles of $1 < \pt < 10\GeV$ and $\abs{\eta} < 1.4$, the track resolutions are typically 25--90 (45--150)\mum in the transverse (longitudinal) impact parameter \cite{TRK-11-001}.

The hadron forward (HF) calorimeter uses steel as an absorber and quartz fibers as the sensitive material. The two halves of the HF are located 11.2\unit{m} from the interaction region, one on each end, and together they provide coverage in the range $3.0 < \abs{\eta} < 5.2$.

Events of interest are selected using a two-tiered trigger system~\cite{Khachatryan:2016bia}. During \XeXe operation the first level trigger (L1), composed of custom hardware processors, uses information from the calorimeters to select events at a rate of around 4\unit{kHz} within a time interval of less than 4\mus. The second level, known as the high-level trigger (HLT), consists of a farm of processors running a version of the full event reconstruction software optimized for fast processing, and reduces the event rate to around 2\unit{kHz} before data storage.

A more detailed description of the CMS detector, together with a definition of the coordinate system used and the relevant kinematic variables, can be found in Ref.~\cite{Chatrchyan:2008zzk}.

\section{Event samples and selections}
\label{sec:evtSelection}
This measurement uses \XeXe data collected at $\sqrtsNN=5.44\TeV$ in 2017. During the six-hour data-taking period approximately 19 million minimum-bias (MB) events were gathered, corresponding to an integrated luminosity of 3.42\mubinv.  Events containing multiple \XeXe collisions have a negligible effect on the measurement, as the average number of interactions per bunch crossing was less than 0.018.  Events selected by the L1 trigger system were required to have a signal above the noise threshold in at least one of the two HF calorimeters.  The HLT chose events having an energy deposit above approximately 1~\GeV in the HF, as well as having at least one group of three pixel hits that is compatible with the trajectory of a charged particle originating from the luminous region.  Every event passing these MB trigger conditions was recorded.

Samples of simulated \XeXe Monte Carlo (MC) events are used to evaluate the detector performance and reconstruction efficiencies.  Both MB \textsc{epos}~\cite{Werner:2005jf} tune LHC~\cite{Pierog:2013ria} and \textsc{hydjet} tuned with $\sqrtsNN=5.02\TeV$ \PbPb MB events~\cite{Lokhtin:2005px} are employed.  An additional set of \textsc{hydjet}-embedded \PYTHIA 8.230~\cite{Sjostrand:2007gs} events (MB \textsc{hydjet} events containing an additional hard scattering generated by \PYTHIA tune CUETP8M1~\cite{Khachatryan:2015pea}) is used to examine the reconstruction performance and \pt resolution for high-\pt charged particles.

A heavy ion collision centrality quantifies the amount of overlap between the two colliding ions.  For both data and MC events, the centrality is estimated from the sum of the transverse energy deposited in both HF detectors.  In this work, centrality selections are expressed as percentage ranges of the total hadronic inelastic cross section.  Lower percentiles indicate a larger degree of overlap between the two nuclei.  Thus, the 0--5\% centrality range selects the most head-on \XeXe collisions in the sample.

\begin{table}[b]
\centering
\topcaption{The values of \avgNPart, \avgNColl, \TAA, and their
uncertainties, for $\sqrtsNN=5.44$\TeV \XeXe collisions and 5.02\TeV \PbPb collisions in the
centrality ranges used here.}
\newcolumntype{x}{D{,}{\text{--}}{2.3}}
\newcolumntype{Y}{D{,}{}{4.4}}
\newcolumntype{Z}{D{,}{}{4.4}}
\newcolumntype{y}{D{,}{}{5.5}}
\newcolumntype{z}{D{,}{}{5.5}}
\setlength\extrarowheight{1.5 pt}
\begin{tabular}{xYZcYZcyz}
\hline
 & \multicolumn{2}{c}{\avgNPart}& & \multicolumn{2}{c}{\avgNColl} & & \multicolumn{2}{c}{\TAA [mb$^{-1}$]}  \\\cline{2-3}\cline{5-6}\cline{8-9}
\multicolumn{1}{c}{Centrality} &\multicolumn{1}{c}{\XeXe}&\multicolumn{1}{c}{\PbPb}&&\multicolumn{1}{c}{\XeXe}&\multicolumn{1}{c}{\PbPb}&&\multicolumn{1}{c}{\XeXe}&\multicolumn{1}{c}{\PbPb} \\
 \hline

0,5\%   & 236.1,{\pm1.3}& 384.3,{^{+1.8}_{-2.0}} && 930,{\pm51}    & 1820,{^{+130}_{-140}} && 13.60,{\pm0.74}   & 26.0,{^{+0.5}_{-0.8}}      \\
5,10\%  & 206.3,{\pm1.7}&333.3,{^{+3.0}_{-3.2}}&& 732,{\pm44}    & 1430,{^{+100}_{-110}} && 10.70,{\pm0.65}   & 20.5,{^{+0.4}_{-0.6}}      \\
10,30\% &141.2,{\pm1.8}&226.7,{^{+5.2}_{-5.3}}&& 407,{\pm30}    & 805,{^{+55}_{-58}}    && 5.94,{\pm0.44}    & 11.5,{^{+0.3}_{-0.4}}      \\
30,50\% & 68.5,{\pm2.2} &109.2,{^{+4.3}_{-4.2}}&& 135,{\pm15}   & 267,{^{+20}_{-20}}    && 1.97,{\pm0.22}    & 3.82,{^{+0.21}_{-0.21}}    \\
50,70\% & 27.2,{\pm1.6} &42.2,{^{+3.0}_{-2.9}}&& 35.3,{\pm4.8} & 65.4,{^{+7.0}_{-6.6}} && 0.517,{\pm0.071}  & 0.934,{^{+0.096}_{-0.089}}\\
70,80\% & 10.55,{\pm0.78}&\multicolumn{1}{c}{\NA}&& 9.8,{\pm1.4}&\multicolumn{1}{c}{ \NA} && 0.143,{\pm0.020}  & \multicolumn{1}{c}{\NA}\\
70,90\% & \multicolumn{1}{c}{\NA}&11.1,{^{+1.3}_{-1.2}}&&\multicolumn{1}{c}{ \NA}& 10.7,{^{+1.7}_{-1.5}} && \multicolumn{1}{c}{\NA}  & 0.152,{^{+0.024}_{-0.021}}\\
0,10\%  & 221.2,{\pm1.5} &358.8,{^{+2.4}_{-2.6}}&& 831,{\pm47}  & 1630,{^{+120}_{-120}} && 12.10,{\pm0.69}   & 23.2,{^{+0.4}_{-0.7}}     \\
\hline
\end{tabular}
\label{tab:TAA}
\end{table}

An event centrality is closely related to the number of participating nucleons, $N_{\text{part}}$, and the number of binary nucleon-nucleon collisions, $N_{\text{coll}}$, in the event.  The \avgNPart, \avgNColl, and corresponding \TAA for a given centrality range are calculated with a Glauber model of the nucleons contained in each ion~\cite{Miller:2007ri}.  For the purposes of this model, the nucleon-nucleon inelastic cross section $\sigma^{\text{inel}}_{\text{NN}}$ is taken as $68.4\pm0.5\unit{mb}$~\cite{Loizides:2017ack}.  The nuclear radius and skin depth are set as $5.36\pm0.1\unit{fm}$ and $0.59\pm0.07\unit{fm}$, respectively~\cite{Loizides:2014vua}. Additionally, the nuclear deformation parameter of the xenon nucleus is taken to be $\beta_{2} = 0.18\pm0.02$~\cite{Acharya:2018hhy}.   Simulated \textsc{epos} events are used to account for bin-to-bin smearing in centrality caused by fluctuations and the energy resolution of the HF calorimeters~\cite{Miller:2007ri}.  The resulting values and uncertainties are given in Table~\ref{tab:TAA} for XeXe collisions at $\sqrtsNN=5.44\TeV$.  For the purpose of calculating $\rxepb$, the same quantities in \PbPb collisions at 5.02\TeV are also given.  The procedure for calculating the \PbPb values is described in Ref.~\cite{Khachatryan:2016odn}.  The uncertainties in the PbPb values include a component related to the uncertainty in the PbPb event selection efficiency.  However, the effect of the XeXe event selection efficiency uncertainty is much larger than in \PbPb collisions.  Therefore, this component is not propagated to the uncertainty in the XeXe values and is accounted for with a separate systematic uncertainty.  In this paper, the definition of \raa containing \TAA, given in Eq.~(\ref{eqn:def_raa2}), is used.

In the offline analysis, events are required to have a reconstructed primary vertex that is formed from at least two tracks and is within $15\unit{cm}$ of the detector center.  This rejects background processes such as beam-gas collisions.  The events must also have at least three detector elements containing energy deposits of at least 3\GeV in each of the two HF subdetectors.  Finally, at least 25\% of the tracks in an event must pass a track-quality selection~\cite{TRK-11-001}. These conditions, along with the MB trigger requirements, are estimated to select $(95\pm3)\%$ of the total inelastic cross section.  This efficiency also includes potential contributions from ultraperipheral electromagnetic interactions contaminating the selected sample and was calculated using samples of \textsc{epos}, \textsc{hydjet}, and \textsc{starlight} v2.2~\cite{Klein:2016yzr}.  In the 0--80\% centrality range used for this analysis, the event selection is fully efficient and any remaining electromagnetic contamination is negligible.

\section{Track reconstruction and corrections}
\label{sec:tracking}
 The spectra measured here are for primary charged particles, defined as having an average proper lifetime greater than $1\unit{cm}$. Daughters originating from secondary decays are not considered primary unless the mother particle has an average proper lifetime under $1\unit{cm}$.  The rate at which these nonprimary tracks contaminate the sample is estimated to be less than $0.3\%$.  Particles coming from interactions with detector components are not included in the primary-particle definition.

Tracks and primary vertices are reconstructed using the procedures described in Ref.~\cite{TRK-11-001}. Small modifications to these algorithms are made to facilitate the reconstruction of \XeXe events having large track multiplicities.  Tracks are required to be in the range $\abs{\eta}<1$. Poor-quality tracks are removed from the sample by applying strict track selections identical to the ones described for $\PbPb$ collisions in Ref.~\cite{Khachatryan:2016odn}.  Notably, these selections require each track with $\pt>20\GeV$ to be associated with a calorimeter energy deposit~\cite{Sirunyan:2017ulk} of at least half the track's momentum.  They also reject tracks having a significance of the distance of closest approach (DCA) to the primary vertex in the $x$-$y$ plane that is greater than 3 standard deviations.

\begin{figure}
  \centering
    \includegraphics[width=0.6\textwidth]{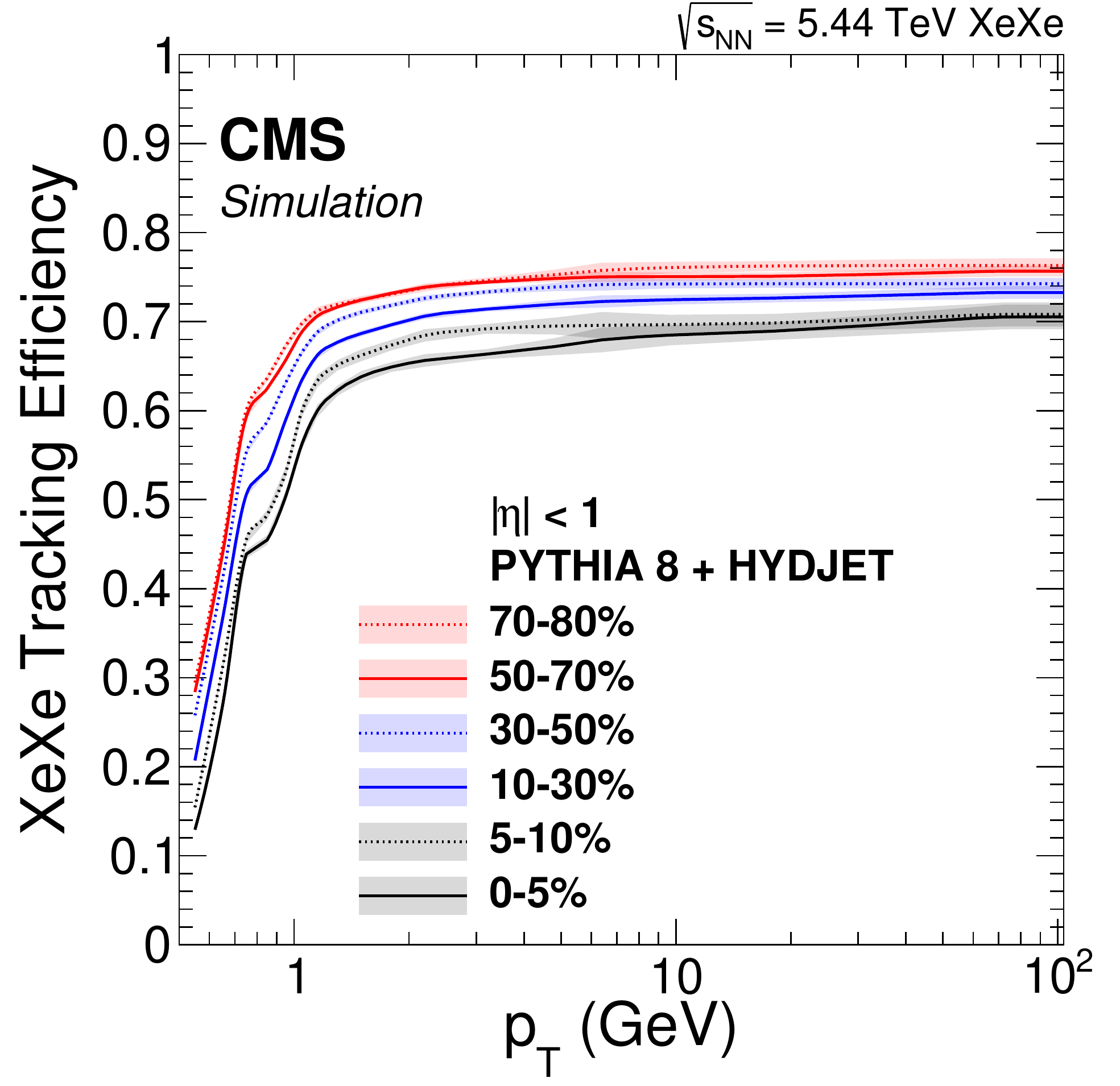}
  \caption{The \XeXe tracking efficiency for six centrality selections.  The tracking efficiency at low-\pt values decreases because of the strict track quality requirements used.  Above $\pt=3\GeV$ the efficiency for central events is rather flat around 70\%.  The shaded bands show the statistical uncertainties.}
  \label{fig:Eff}
\end{figure}

The tracking performance is evaluated using simulated \textsc{hydjet}-embedded \PYTHIA events and is found to be similar to the performance in \PbPb collisions having similar detector occupancy.  The track \pt resolution is ${<}1.5\%$ for the full \pt range of this study.  The tracking efficiency, defined as the fraction of primary charged particles successfully reconstructed after track quality selections, is shown in Fig.~\ref{fig:Eff}.  The shaded bands around each line show statistical uncertainties.  The efficiency has a fairly constant value around 70\% (76\%) in the range $3<\pt<100\GeV$ for central (peripheral) events.  Because of the stringent track selection criteria, the efficiency decreases to a value of $13\%$ at $\pt=0.5\GeV$ in the 0--5\% centrality range, and to $30\%$ in the 70--80\% centrality range.  The rate at which erroneous tracks not associated with a charged particle are generated, or the misreconstruction rate, is less than $1\%$ for most of the \pt range studied.  However, it does increase quickly for tracks having $\pt<0.7\GeV$ in the 0--5\% centrality range, reaching a maximum value of 34\% at $\pt=0.5\GeV$.  The effects of tracking inefficiency, misreconstruction, and nonprimary contamination are all corrected for by applying a weight to each track.  This correction is parameterized as a function of the track \pt and event centrality.

The tracking efficiency for a charged particle at a given \pt depends on its species.  Additionally, some charged particles, notably the strange baryons, are more likely to decay into secondary particles which then contaminate the sample.  These effects lead to a model dependence of the total tracking correction, because different MC event generators predict dissimilar relative fractions of each type of charged particle.  Notably, \PYTHIA tends to underpredict strange hadron production in $\Pp\Pp$ collisions~\cite{Acharya:2018qsh}, while \textsc{epos} is found to overestimate the production of many strange hadrons in central \PbPb collisions~\cite{ABELEV:2013zaa}.  Thus, the fraction of strange baryons in data is expected to be bounded by that of \textsc{epos} and of the embedded particles in a \textsc{hydjet}-embedded \PYTHIA sample.  Following the procedure detailed in Ref.~\cite{Khachatryan:2016odn}, a working point is chosen that lies halfway between the tracking corrections produced by these two generators.  The deviation between the estimated tracking corrections from the two generators reaches a maximum of 8\% around $\pt=4\GeV$ but is less than 3\% for $\pt>10\GeV$.

\section{Reference spectrum}
\label{sec:ppRef}

\begin{figure}
  \centering
    \includegraphics[width=0.65\textwidth]{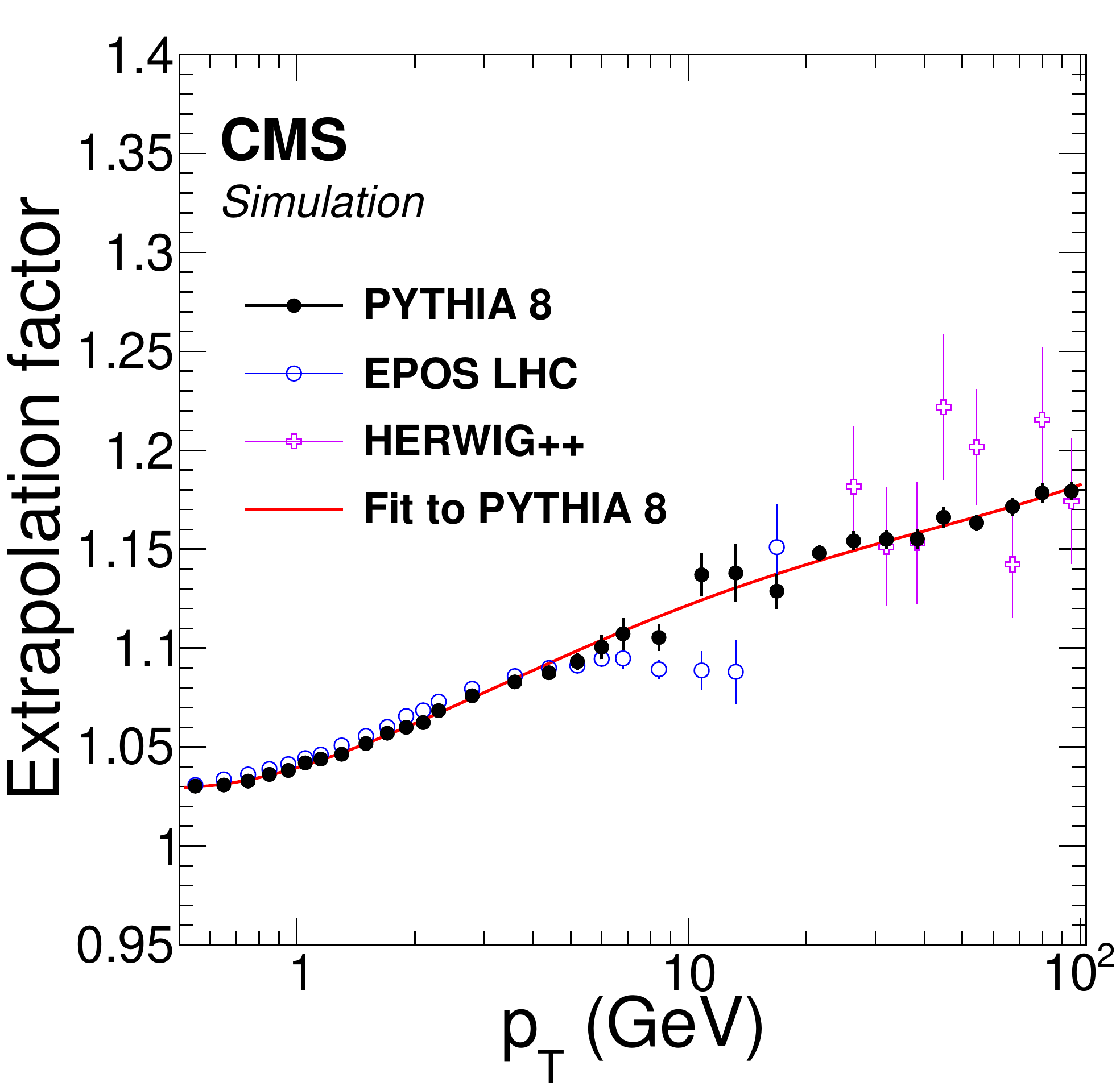}
  \caption{The ratio of charged-particle spectra in $\Pp\Pp$ collisions at $\sqrts=5.44$ and 5.02\TeV for three different MC generators.  A fit to the \PYTHIA ratio is shown by the red line.}
  \label{fig:ppRefExtrap}
\end{figure}

A reference spectrum from $\Pp\Pp$ collisions at an appropriate center-of-mass energy is required to construct \raa.  Although no measurements exist at $\sqrts=5.44\TeV$, the CMS Collaboration has measured \pt spectra for collisions at $\sqrts=5.02\TeV$~\cite{Khachatryan:2016odn} and 7\TeV~\cite{Chatrchyan:2011av}.  An MC-based extrapolation procedure is applied to the 5.02\TeV spectrum because of its close proximity in energy to 5.44\TeV. The $\Pp\Pp$ reference cross section used for the \raaStar calculation is
\begin{linenomath}
\begin{equation}
\label{eq:extrapolation}
\Bigg(\dd{\sigma_{\text{5.44}}^{\Pp\Pp}}{\pt}\Bigg)_{\text{Extrap.}} =  \Bigg(\dd{\sigma_{\text{5.44}}^{\Pp\Pp}}{\pt} \Bigg/ \dd{\sigma_{\text{5.02}}^{\Pp\Pp}}{\pt}\Bigg)_{\text{MC}} \Bigg(\dd{\sigma_{\text{5.02}}^{\Pp\Pp}}{\pt}\Bigg)_{\text{Data}}.
\end{equation}
\end{linenomath}
For most of the \pt range studied here, the charged-particle spectra for $\Pp\Pp$ collisions produced by \PYTHIA 8.223 tune CUETP8M1 were found to match data at $\sqrts=5.02$ and 7\TeV within the experimental uncertainties.  Differences between the data and simulation for $\pt<1\GeV$ and around $\pt=10\GeV$ are similar at both center-of-mass energies and are expected to largely cancel in a ratio.  Therefore, this generator is used for the reference reported here.  The extrapolation factor is extracted by fitting a polynomial of the form $a_0+a_1x+a_2x^2+a_3x^3+a_4x^4$, with $x=\ln(\pt/1\GeV)$, to the ratio of spectra at the two different center-of-mass energies.  The fit parameters are $a_0=1.04$, $a_1=2.56 \times 10^{-2}$, $a_2=1.27 \times 10^{-2}$, $a_3=-4.72 \times 10^{-3}$, and $a_4=4.80 \times 10^{-4}$. This functional form is chosen to give a good empirical description of the simulated data, as seen in Fig.~\ref{fig:ppRefExtrap}, and is not guaranteed to be valid outside the range $0.5<\pt<100\GeV$.  The extrapolation factor spans the range from 1.03 at $\pt=0.5\GeV$ to 1.18 at $\pt=100\GeV$.  For most of this \pt range, the fit's statistical uncertainty is smaller than the thickness of the red line in Fig.~\ref{fig:ppRefExtrap}.  The extrapolation procedure is checked at low-\pt using \textsc{epos} tune LHC, which is found to be within 1\% of \PYTHIA until around $\pt=10\GeV$.  At higher \pt, a fit to \textsc{herwig++}\cite{Bahr:2008pv} tune EE5C~\cite{Khachatryan:2015pea} deviates from the \PYTHIA result by no more than 2\%.  Other functional forms including sigmoid functions and ratios of Tsallis distributions~\cite{Wong:2012zr} are found to agree with the nominal fit to within 1\%.

Alternative methods of calculating a reference spectrum were attempted.  A similar extrapolation starting from data at $\sqrts=7\TeV$ is found to yield a reference spectrum within 5\% of the one constructed using 5.02\TeV data.  This difference is well within the experimental uncertainties of the 5.02 and 7\TeV data.  The spectra produced by ``relative placement'' and $x_{T}$ interpolation procedures~\cite{Khachatryan:2015xaa} are tightly constrained by the existing 5.02\TeV measurement and are within 2\% of the extrapolated reference cross section used here.

\label{sec:uncertainties}
\begin{table}[tbh]
\centering
\topcaption{The systematic uncertainties related to the measurements reported here. The values quoted cover the centrality and \pt dependence of each uncertainty. They are separated into normalization uncertainties and all other systematic uncertainties.}
\begin{tabular}{lccc}
 \hline
  Sources & \multicolumn{3}{c}{Uncertainty [\%]} \\\cline{2-4}
  & \XeXe Spectra & \raaStar & \rxepb \\
 \hline
  Fraction of misreconstructed tracks & 0.1--16 & 0.1--16 & 0.1--5 \\
  Particle species composition & 0.5--8 & 0.5--8 & 1--8 \\
  Track selection & 3--6 & 3--6 & 5--7 \\
  MC/data tracking efficiency difference & 5 & 2.0--6.4 & \NA \\
  Tracking corrections & 0.5--2 & 0.5--2 & 1--5 \\
  \pt resolution & 0.5 & 0.5 & \NA \\
  Extrapolated $\Pp\Pp$ reference& \NA & 4--9 & \NA \\
  Trigger combination& \NA & \NA & 1 \\[\cmsTabSkip]
  Combined uncertainty & 7--18 & 6--18 & 6--11 \\[\cmsTabSkip]
  XeXe event selection efficiency& 0.3--26 & 0.3--26 & 0.3--26 \\
  Glauber model uncertainty (\TAA) &\NA& 5--14 & 6--21 \\
  $\Pp\Pp$ reference luminosity & \NA & 2.3 & \NA \\[\cmsTabSkip]
 Combined normalization uncertainty & 0.3--26 & 6--30 & 6--33 \\
\end{tabular}
\label{tab:systUncerts}
\end{table}

\section{Systematic uncertainties}

A breakdown of the systematic uncertainties related to measurements of the \XeXe charged-particle \pt spectra, \raaStar and \rxepb is given in Table~\ref{tab:systUncerts}.  Systematic uncertainties that are fully correlated between points in a given centrality range are grouped together as normalization uncertainties and are not combined with other uncertainties.  The ranges reported cover the span of each uncertainty across the \pt and centrality range of the measurement.  A detailed discussion of each component of the systematic uncertainty is given below.   References to the uncertainties in \PbPb and $\Pp\Pp$ collisions concern the measurements described in Ref.~\cite{Khachatryan:2016odn}.

\textbullet~Fraction of misreconstructed tracks. The misreconstruction rate is evaluated in simulated events. To account for potential deviations from this value in data, the distribution of the significance of the tracks' DCA to the primary vertex in the $x$-$y$ plane is examined.  The relative contribution of misreconstructed tracks to this distribution is scaled in simulated events to match data in a sideband region having a DCA significance between 25 and 30 standard deviations.  Tracks in this region are almost entirely misreconstructed tracks, and therefore give an estimate of the difference in the misreconstruction effect between data and simulation.  After this scaling procedure, the relative change of the misreconstruction rate in the signal region (less than 3 standard deviations) is taken as the systematic uncertainty.  This is ${<}~2\%$ for most of the data in this analysis.  For tracks having $\pt<0.7\GeV$ in central events, however, it quickly grows to a value of 16\%.

\textbullet~Particle species composition.  The correction applied to account for the model-dependence of the tracking correction assumes the particle composition of data lies somewhere between \PYTHIA and \textsc{epos}.  To cover the range spanned by both of these models, the difference between the two tracking corrections produced by these models is taken as an approximate estimate of the uncertainty.  This uncertainty strongly peaks around 4\GeV, where the difference in particle composition is the largest for the two generators.  At $\pt>10\GeV$, where the two generators converge, a systematic uncertainty of 3\% is assigned.  No cancellation of this uncertainty is assumed for \raaStar.  The uncertainties are correlated in \PbPb and \XeXe collisions and are partially canceled for \rxepb.

\textbullet~Track selection. Differences between data and MC track distributions cause the same track selections to remove slightly different numbers of particles.  The sensitivity of the analysis to this effect is checked by varying the strictness of the track selection criteria.  An uncertainty of 6\% is assigned for this effect under $\pt=20\GeV$.  For higher \pt values the uncertainty is only 3\%.  This uncertainty is conservatively assumed to not cancel in the ratios measured, and a similar uncertainty for \PbPb collisions is included for \rxepb.

\textbullet~MC/data tracking efficiency difference. An uncertainty of 5\% is assigned for additional differences in the tracking efficiency not related to the particle fractions modeled in MC events.  These differences could be related to small variations in the detector conditions or slight inaccuracies in the simulation of the detector.  This uncertainty is estimated using measurements of the relative tracking efficiency in decays of $\PD^{*}$ mesons in pp collisions, along with studies of the relative tracking efficiency's occupancy-dependence in \PbPb collisions.    For \raaStar this systematic uncertainty is conservatively assumed to cancel as much as it did for previous analyses in \PbPb collisions~\cite{Khachatryan:2016odn}, giving an uncertainty of 2.0 (6.4)\% for peripheral (central) events.  This uncertainty largely cancels in \rxepb, where the occupancies of the two systems in the ratio are more similar than in \raaStar.

\textbullet~Tracking corrections. The statistical uncertainty in the tracking corrections, caused by the finite size of the XeXe MC samples used, is accounted for as a systematic uncertainty in the final results.  This uncertainty is between 0.5\% and 2.0\%.  A similar uncertainty covering MC sample size and tracking correction procedures in \PbPb collisions is added in quadrature to this uncertainty for \rxepb.

\textbullet~Transverse momentum resolution. The distortion of the \pt spectra caused by detector resolution was evaluated with simulated events.  A systematic uncertainty of 0.5\% accounts for potential changes in the yield of any given \pt bin.  Because of the similarity in shape of the \XeXe and \PbPb spectra, this uncertainty cancels for \rxepb.

\textbullet~Extrapolated $\Pp\Pp$ reference. The total uncertainty in the extrapolated $\Pp\Pp$ reference cross section at 5.44\TeV is dominated by the 7--10\% uncertainty in the original measurement at 5.02\TeV.  This uncertainty includes a fully correlated 2.3\% uncertainty in the total integrated luminosity~\cite{CMS-PAS-LUM-16-001} that is included as a normalization uncertainty in figures displaying \raaStar.  For the purposes of calculating \raaStar, the MC/data track efficiency difference and \pt resolution components of this uncertainty, which partially cancel with \XeXe uncertainties, are removed from the $\Pp\Pp$ reference data uncertainty and included elsewhere to avoid double counting. An additional 1\% uncertainty is included to account for variations in the functional form used to fit the simulation-based extrapolation factor.

\textbullet~Trigger combination. The \XeXe data used in this analysis were collected with only one MB trigger, so there is no uncertainty related to using multiple triggers to select \XeXe events.  However, the trigger scheme used to measure the \PbPb spectra used in the \rxepb calculation has a 1\% uncertainty associated with it.

\textbullet~XeXe event selection efficiency. The 3\% uncertainty on the total \XeXe event selection efficiency is propagated to the results by repeating the analysis after appropriately varying the centrality calibration.  These variations each cause a shift in the centrality values of the entire data sample, with peripheral centralities being altered significantly more than central ones.  Therefore, this uncertainty is small for central events but grows with the collision centrality.  In the 70--80\% centrality range it reaches values of 26\%.  The uncertainty is fully correlated across all \pt values in a given centrality selection.

\textbullet~Glauber model uncertainty. The uncertainty in \TAA for \XeXe collisions ranges from 5\% to 14\%.  This uncertainty is calculated by propagating uncertainties in the Glauber model's input parameters, which are detailed in Section~\ref{sec:evtSelection}.  The uncertainty in the \XeXe collision event selection efficiency is not included because it is accounted for with a separate systematic uncertainty.  The uncertainty in the quantity $T_{\text{PbPb}}/T_{\text{XeXe}}$, used in \rxepb, is determined by adding in quadrature the relative uncertainties in \TAA for each collision system.

\section{Results}
\label{sec:results}
\begin{figure}
  \centering
    \includegraphics[width=0.9\textwidth]{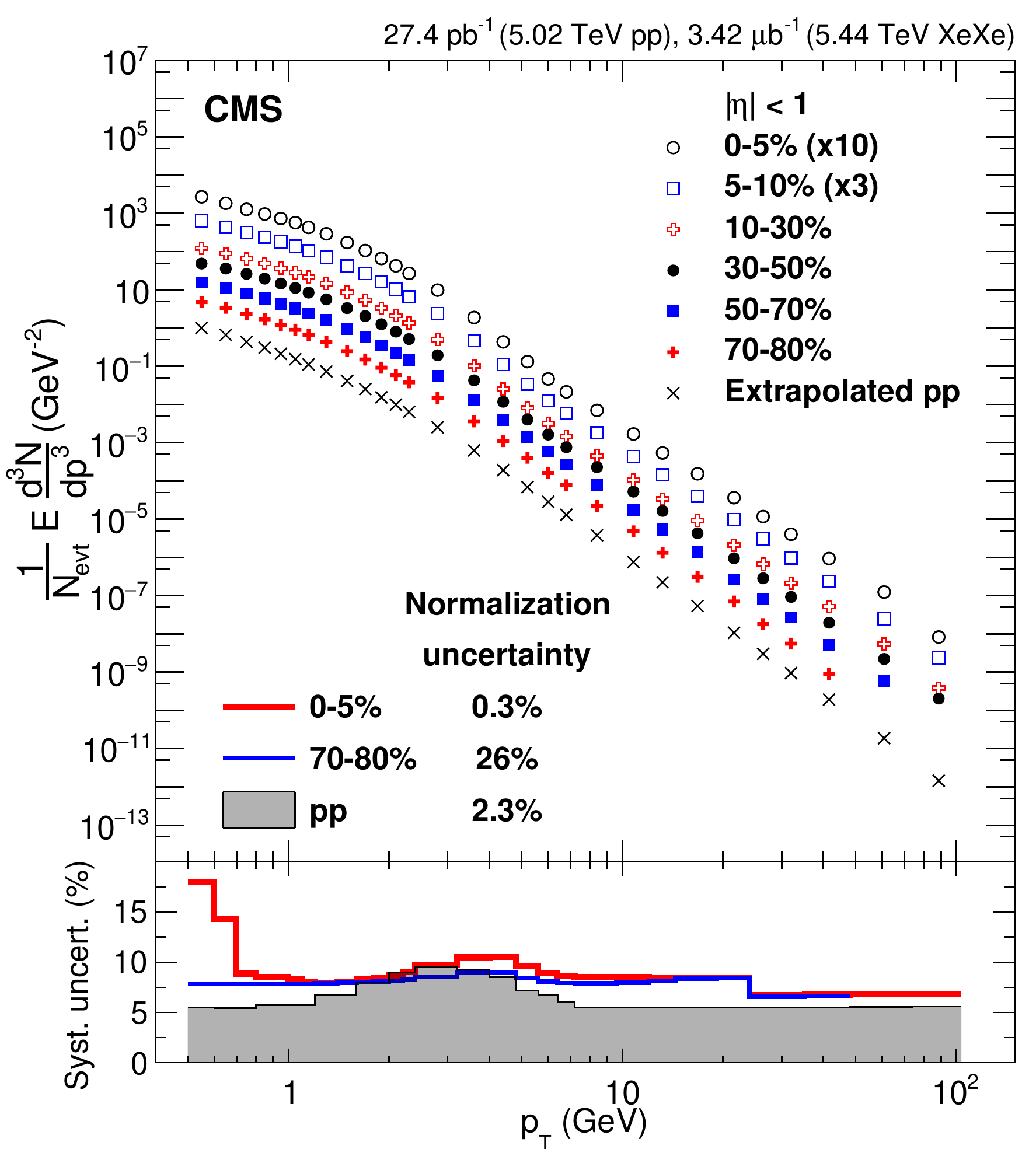}
  \caption{(Upper panel) The charged-particle \pt spectra in six classes of \XeXe centrality and the $\Pp\Pp$ reference spectrum after being extrapolated to $\sqrts=5.44\TeV$.  The statistical uncertainties are smaller than the markers for many of the points.  To facilitate direct comparison, the $\Pp\Pp$ points are converted to per-event yields using a constant factor of $70\unit{mb}$. (Lower panel) The systematic uncertainties for central and peripheral \XeXe collisions, as well as for the $\Pp\Pp$ reference data.}
  \label{fig:XeXeSpectra}
\end{figure}

Charged-particle \pt spectra in \XeXe collisions at $\sqrtsNN=5.44\TeV$ are shown in Fig.~\ref{fig:XeXeSpectra} for six centrality ranges.  The data are reported as per-event invariant differential yields.  To improve visual clarity, the spectra for the 0--5\% and 5--10\% centrality ranges have been scaled by ten and three, respectively.  The extrapolated $\Pp\Pp$ reference data for the same center-of-mass energy is also reported.  The reference used for \raaStar is a differential cross section, but has been converted to a per-event yield using a constant factor of $70\unit{mb}$ to allow for direct comparison in Fig~\ref{fig:XeXeSpectra}.  The data points represent the average charged-particle yield in each \pt bin, not the charged-particle yield at the bin center where the point is placed.  The statistical uncertainties of the measurement are smaller than the markers for most of the data points.  The $\Pp\Pp$ reference spectrum has a shape similar to that of a Tsallis distribution, including a power law behavior at large \pt values.  This is consistent with earlier observations that this functional form is able to describe charged-particle \pt spectra at LHC energies~\cite{Wong:2012zr}.  The lower panel of Fig.~\ref{fig:XeXeSpectra} shows the systematic uncertainties for the most central and peripheral \XeXe collisions, and for the extrapolated $\Pp\Pp$ reference data.  A few values of the systematic uncertainties in the normalization of the spectra are also listed.

\begin{figure}
  \centering
    \includegraphics[width=0.49\textwidth]{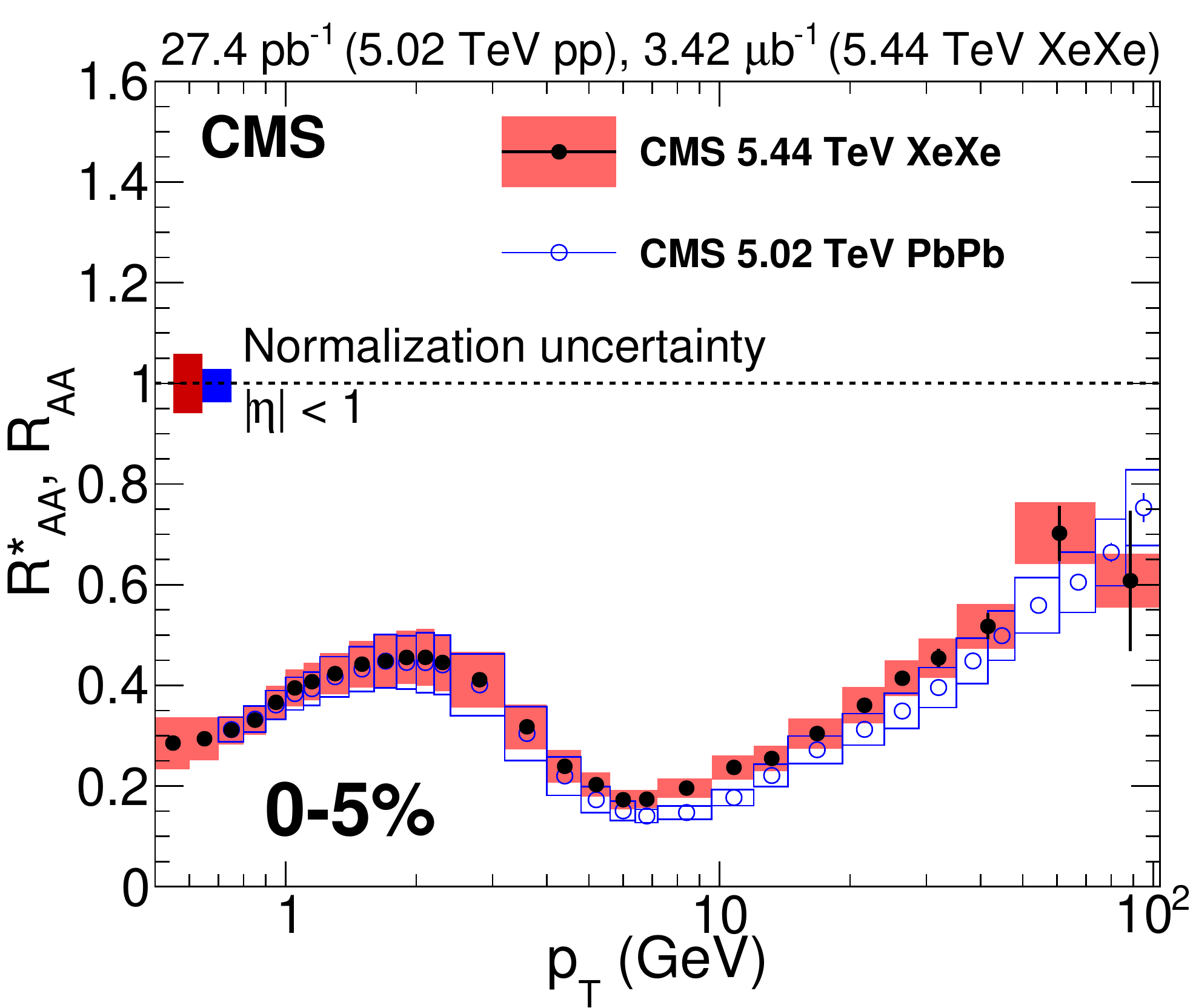}
    \includegraphics[width=0.49\textwidth]{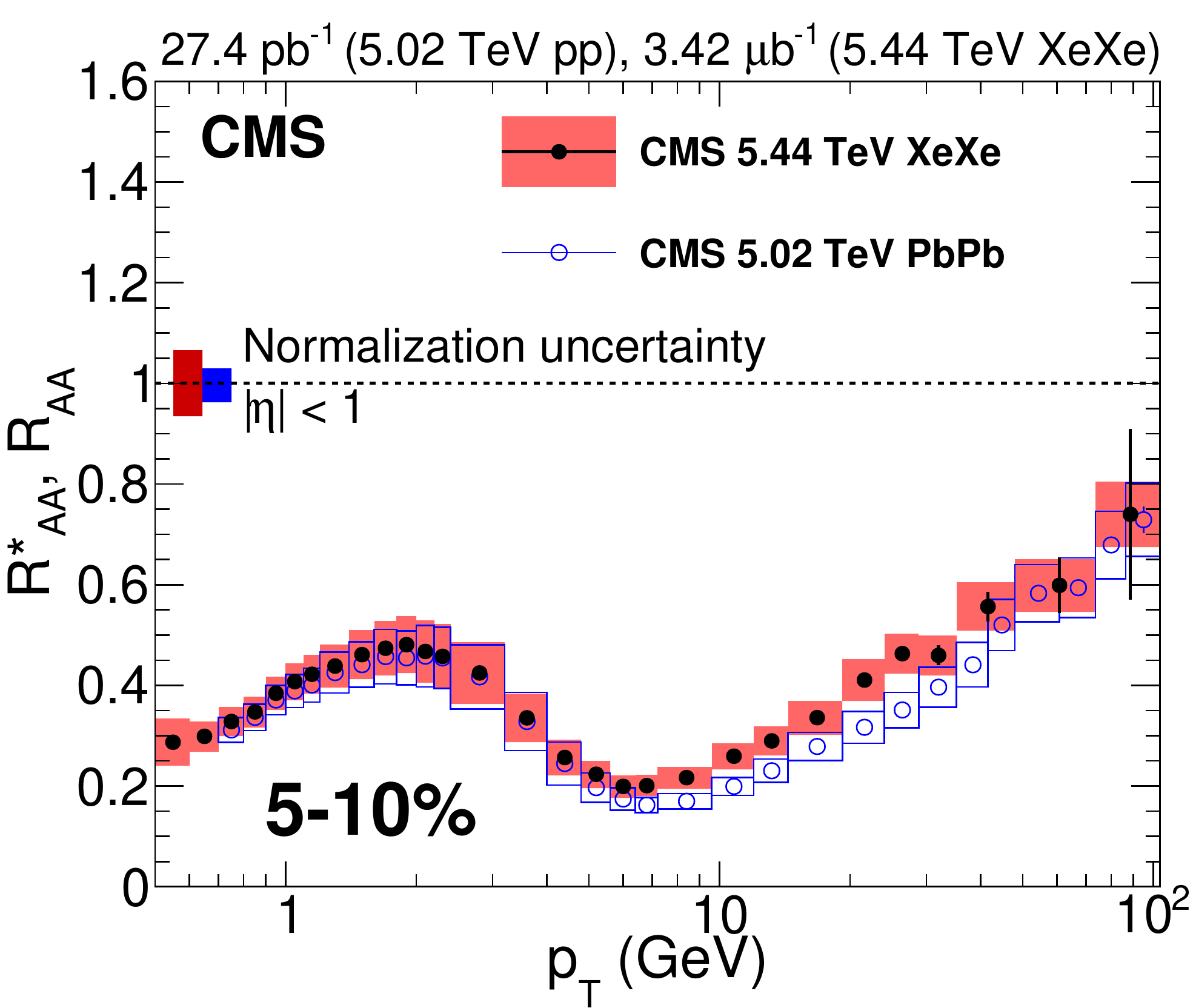}
    \includegraphics[width=0.49\textwidth]{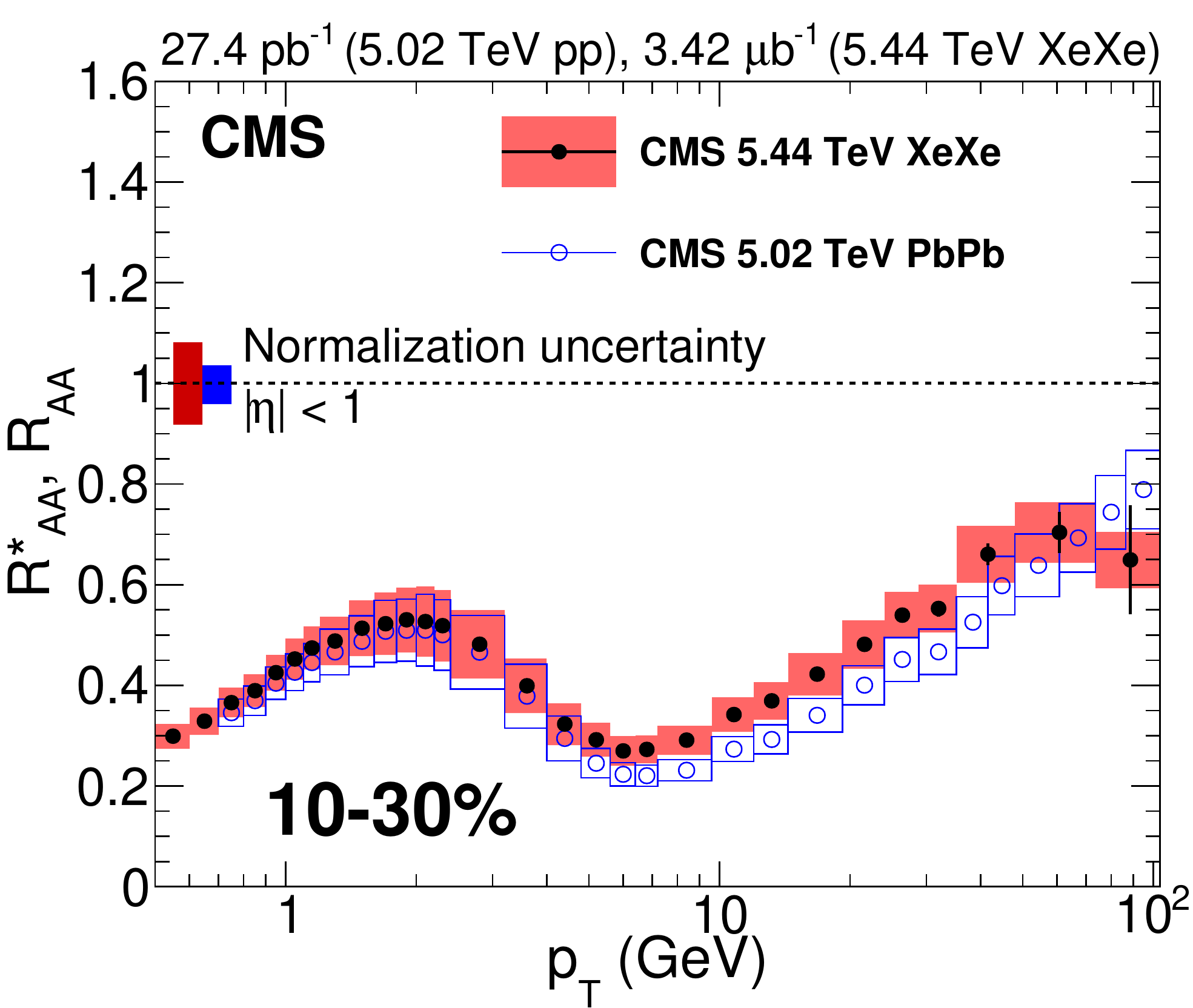}
    \includegraphics[width=0.49\textwidth]{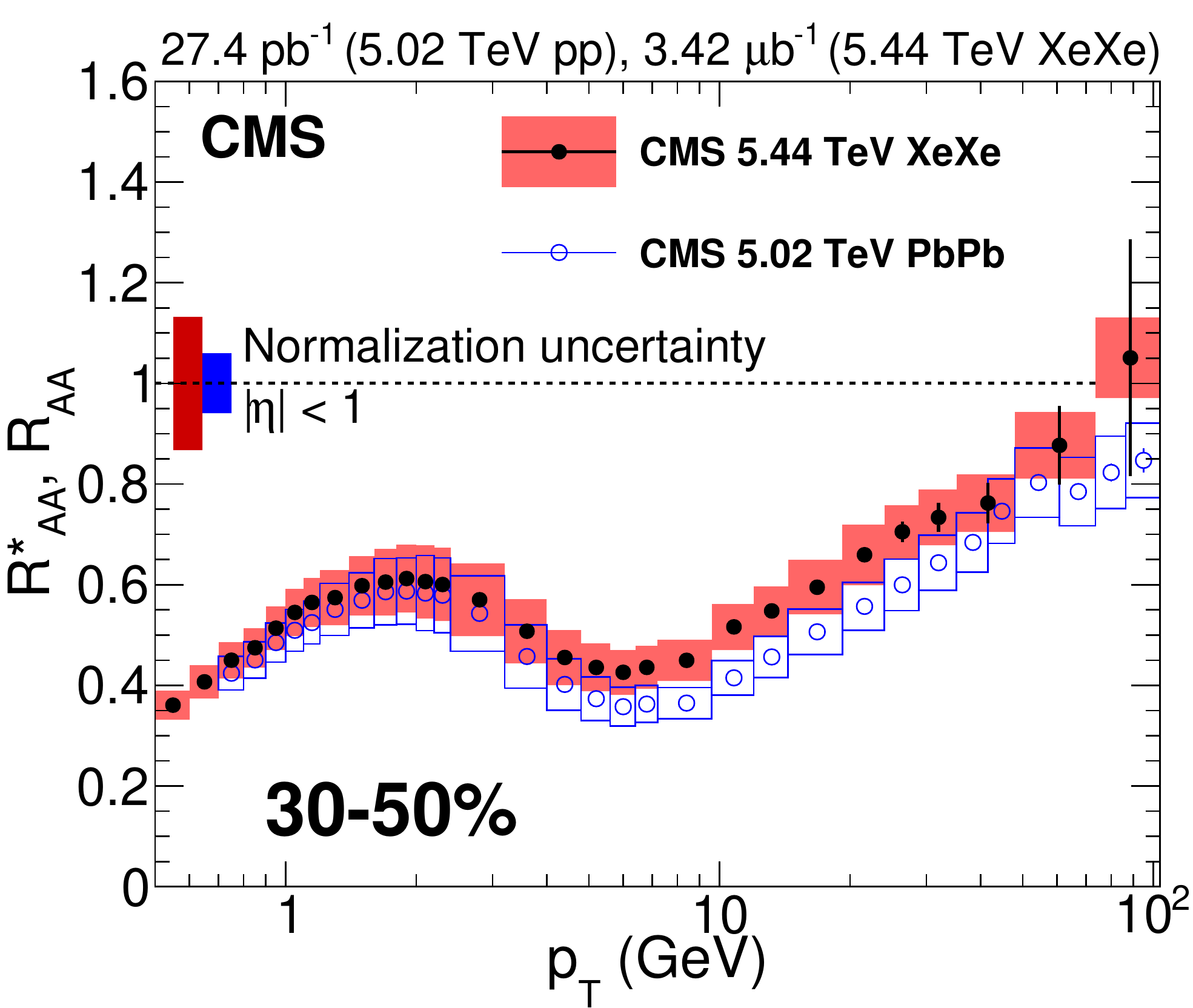}
    \includegraphics[width=0.49\textwidth]{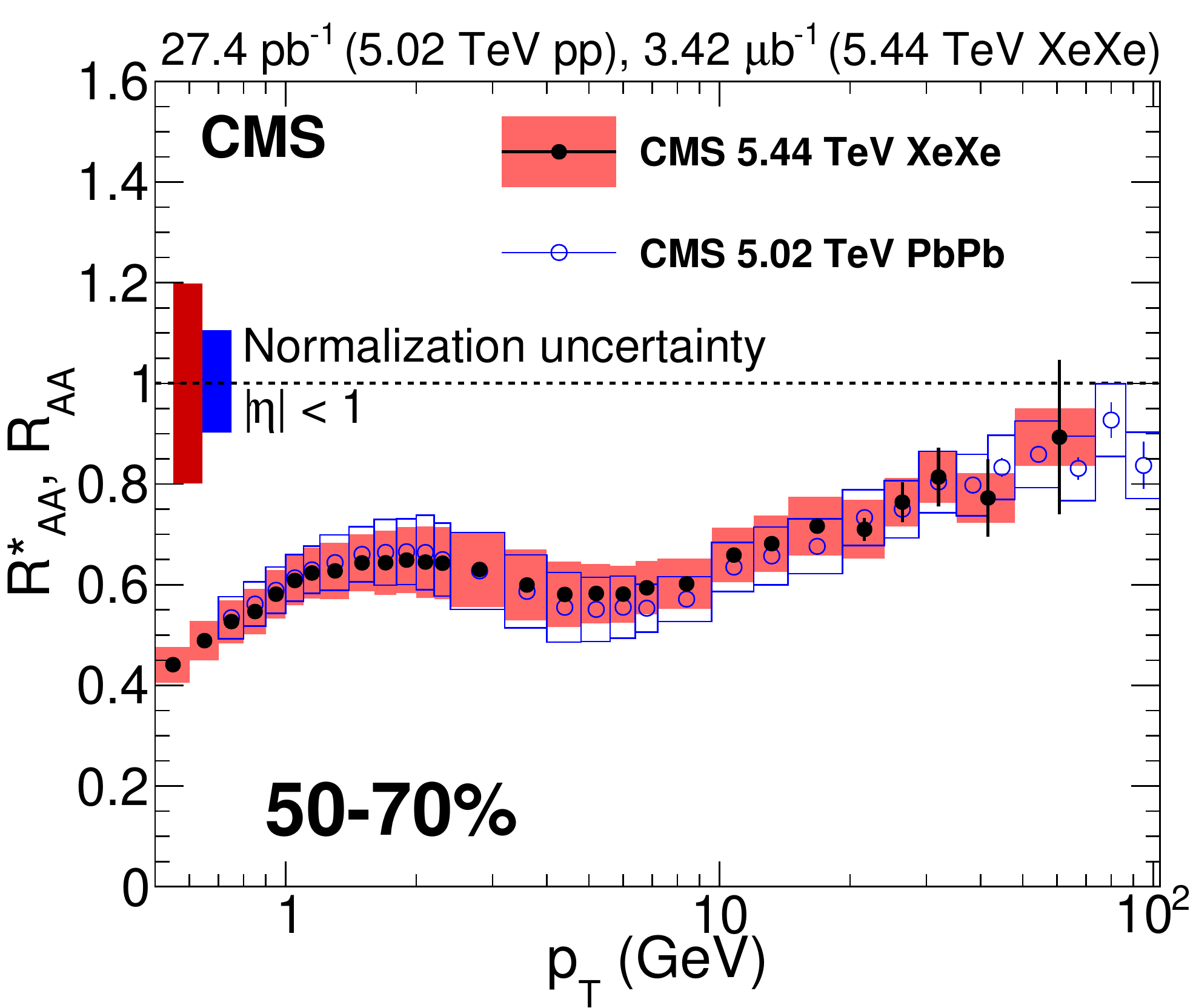}
    \includegraphics[width=0.49\textwidth]{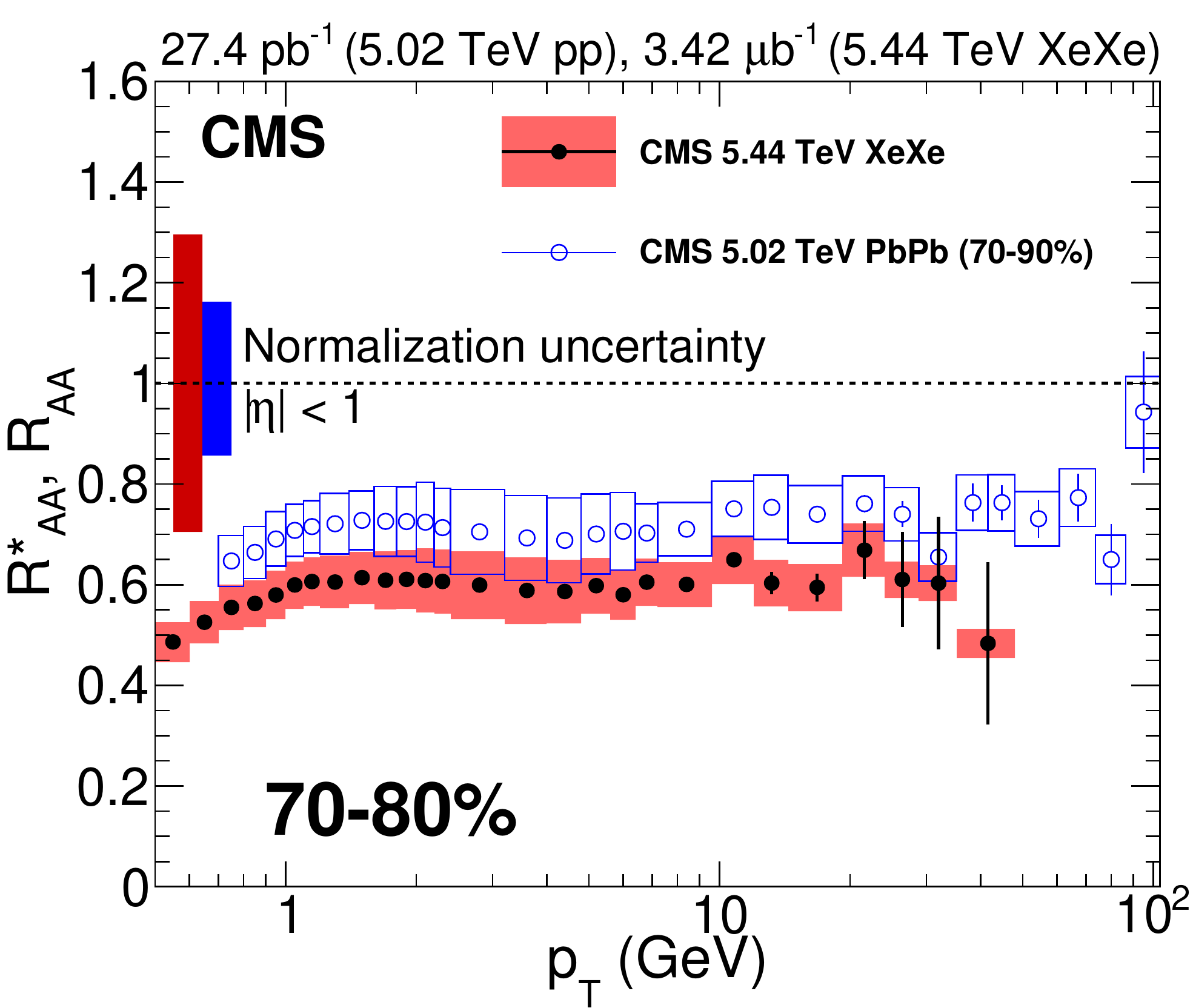}
  \caption{The charged-particle \raaStar for \XeXe collisions at $\sqrtsNN=5.44\TeV$ in six centrality ranges.  A previous measurement of \raa in \PbPb collisions at 5.02\TeV is also shown~\cite{Khachatryan:2016odn}.  The solid pink and open blue boxes represent the systematic uncertainties of the \XeXe and \PbPb data, respectively.}
  \label{fig:RAAResults}
\end{figure}
\begin{figure}
  \centering
    \includegraphics[width=0.49\textwidth]{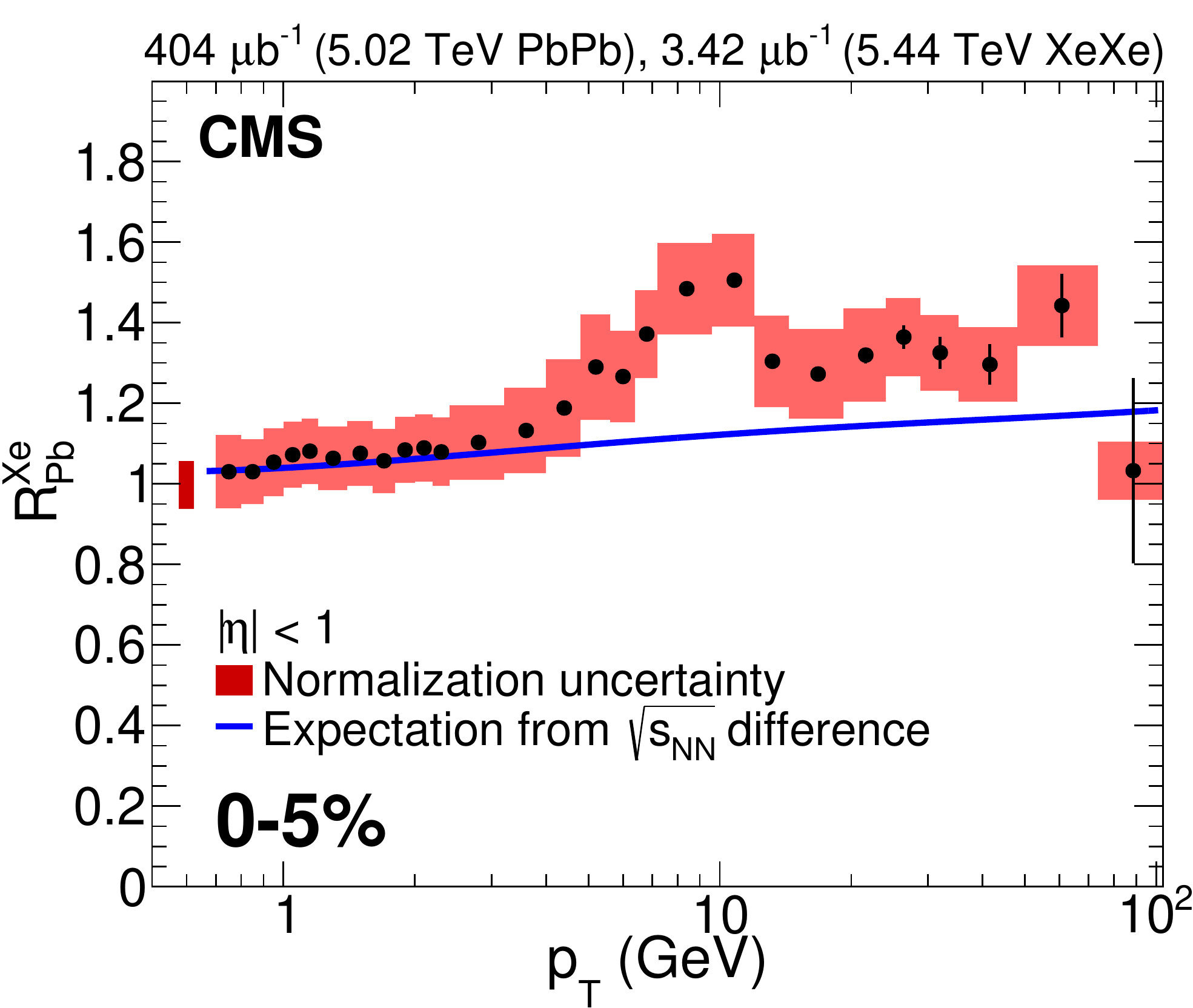}
    \includegraphics[width=0.49\textwidth]{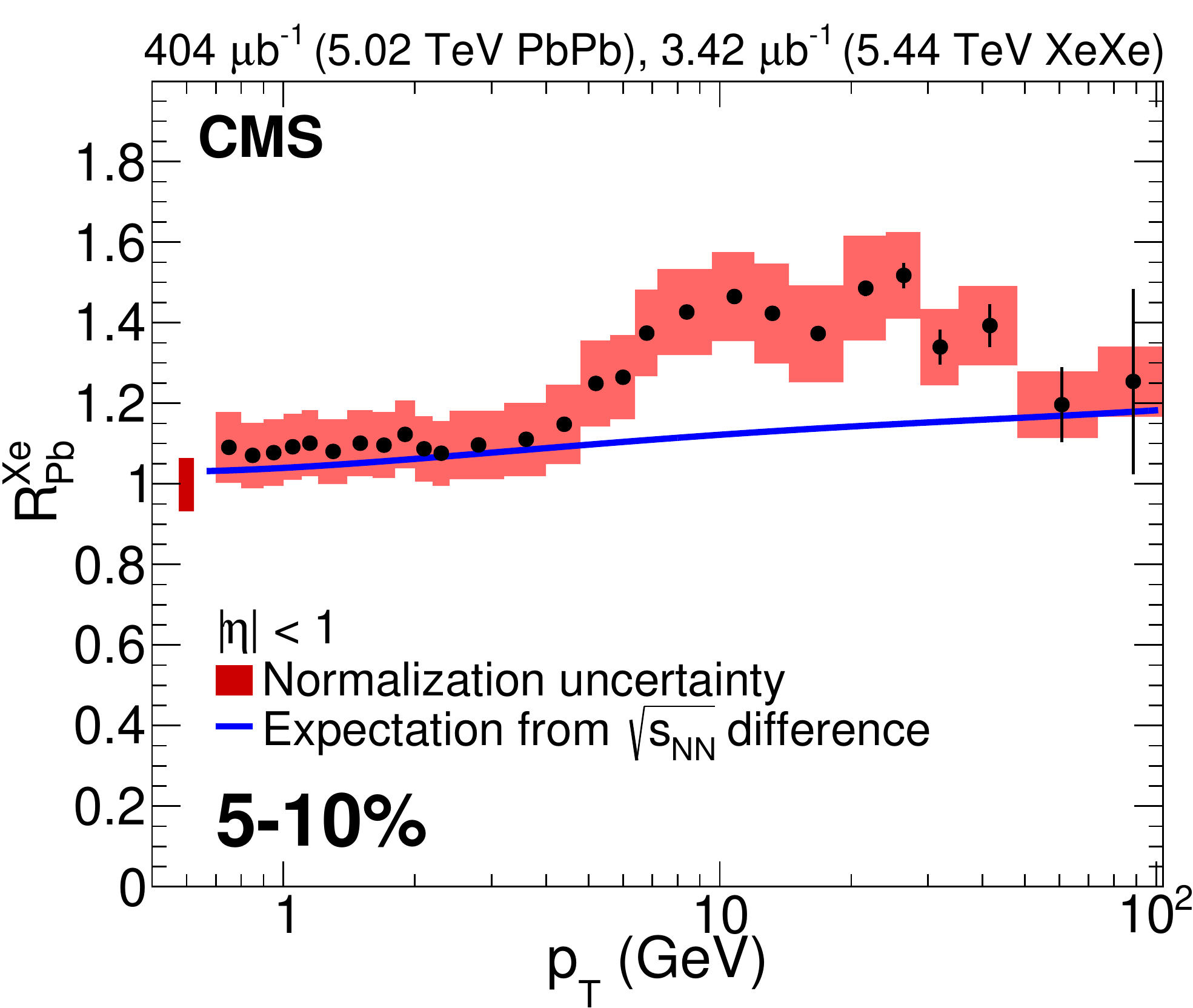}
    \includegraphics[width=0.49\textwidth]{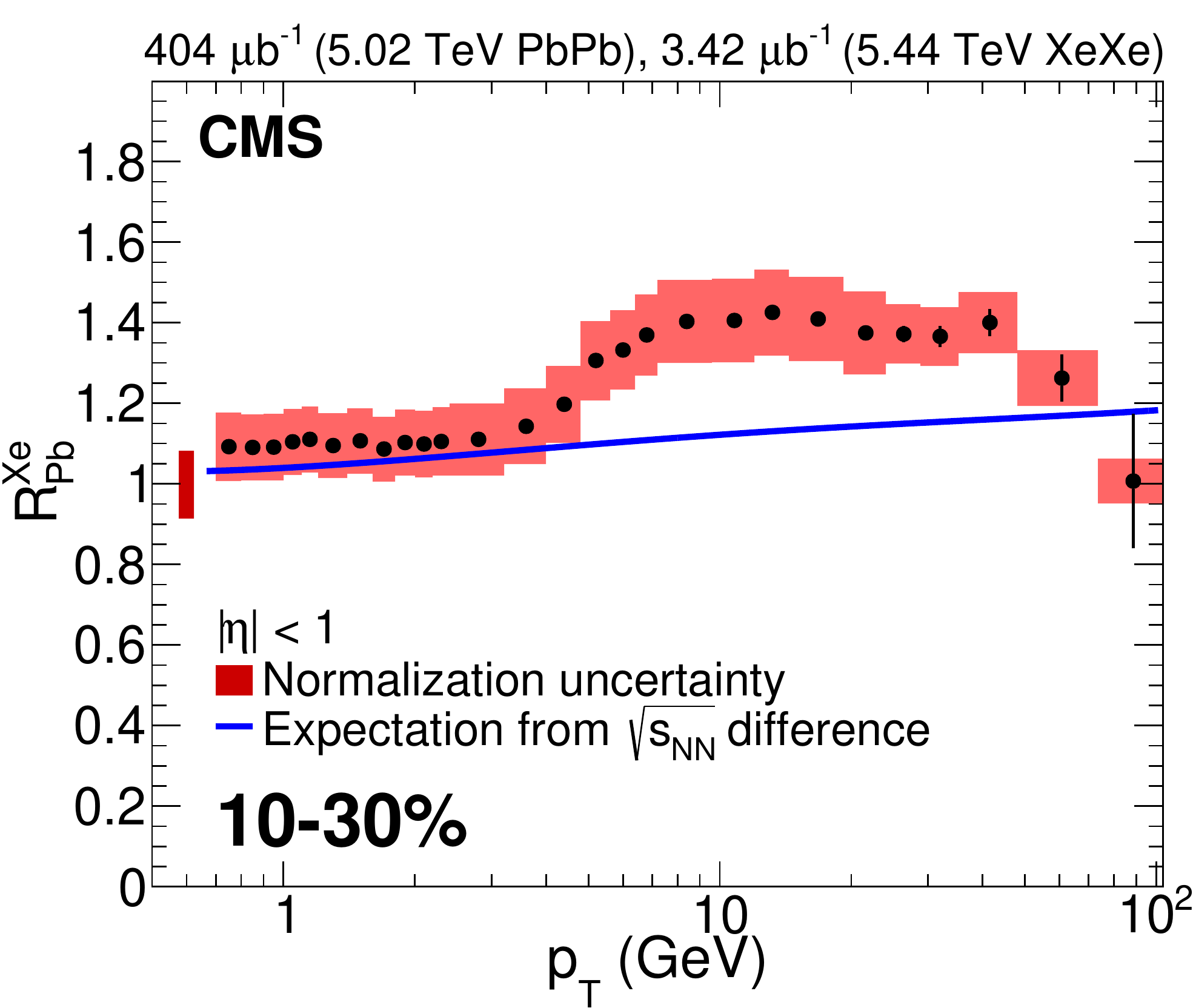}
    \includegraphics[width=0.49\textwidth]{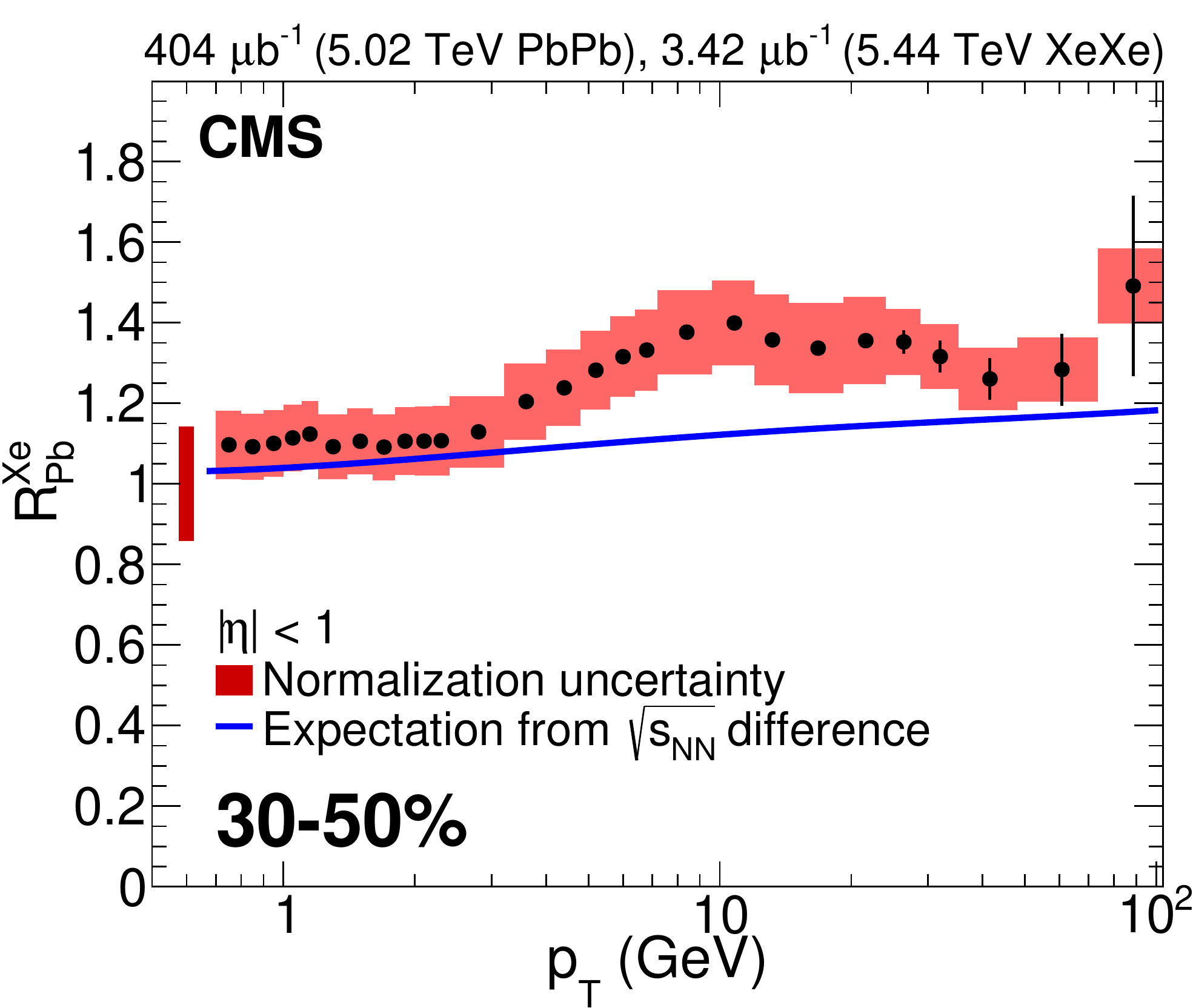}
    \includegraphics[width=0.49\textwidth]{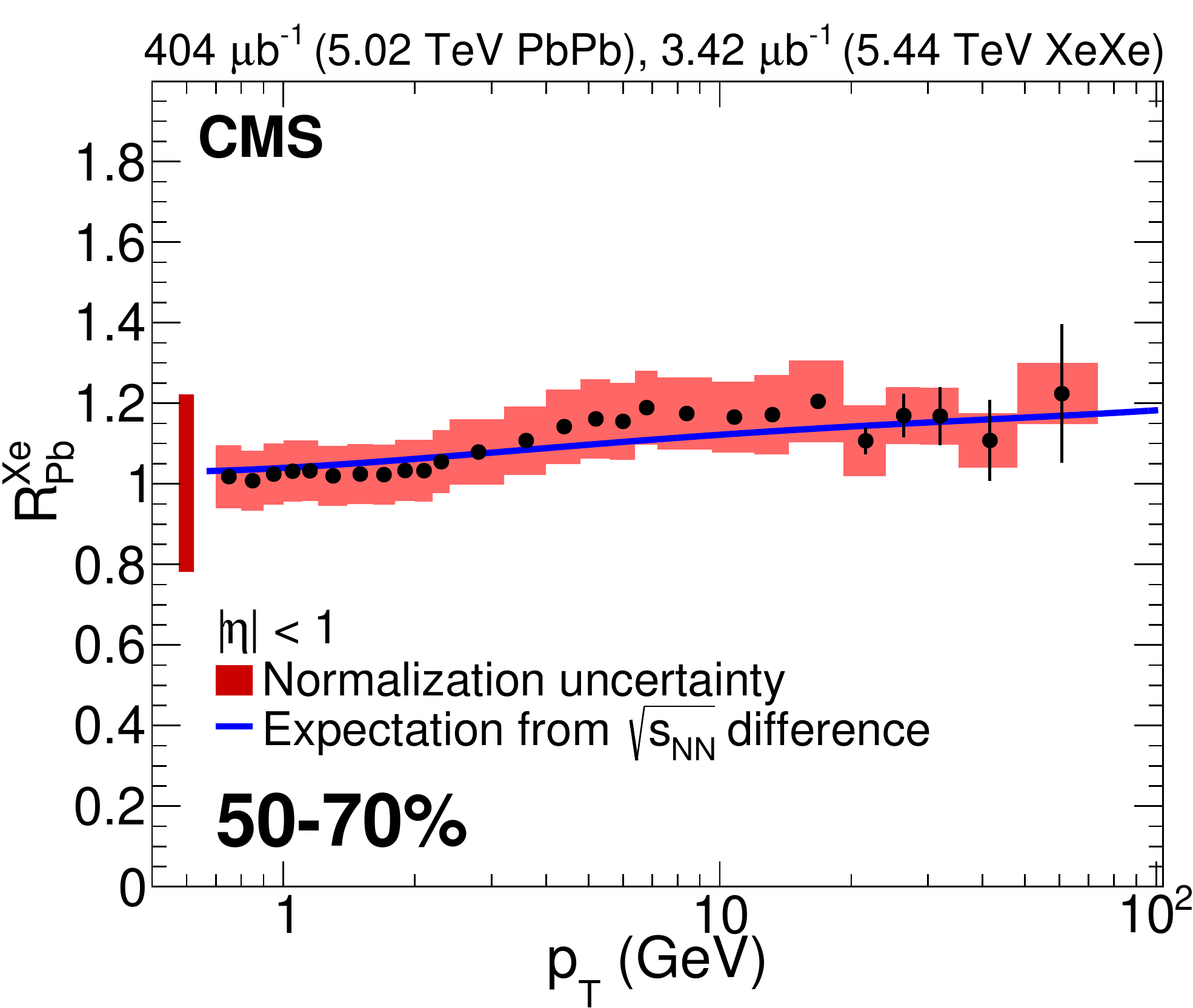}
  \caption{The measurement of \rxepb in five centrality classes using the results of this analysis and data from Ref.~\cite{Khachatryan:2016odn}.  The blue line represents the expected deviation from unity caused by the different center-of-mass energies of the two collision systems.  The solid pink boxes represent the systematic uncertainties.}
  \label{fig:XeXevsPbPb}
\end{figure}

The resulting \raaStar values for primary charged particles in \XeXe collisions are shown in Fig.~\ref{fig:RAAResults}.  The pink boxes represent all systematic uncertainties other than the uncertainty in the overall normalization, which is shown by the dark red box around unity.  The error bars give the statistical uncertainty of the measurement.  For comparison, the \raa in \PbPb collisions at $\sqrtsNN=5.02\TeV$~\cite{Khachatryan:2016odn} is shown by the hollow blue points. The blue boxes represent the systematic uncertainties of the \PbPb data.  The most central events show a strong modification that is most pronounced in the range $5<\pt<30\GeV$.   A similar oscillatory shape is observed in both \XeXe and \PbPb collisions, indicating that hot medium effects seen in \PbPb collisions are also present in \XeXe collisions.  At low \pt, these effects include contributions from the nuclear parton distribution function~\cite{Helenius:2012wd}, radial flow~\cite{Chatrchyan:2012ta}, and the Cronin effect~\cite{Adare:2013esx}. At higher \pt, parton energy loss also becomes a significant effect.  Generally, \raa and \raaStar agree with each other in the range $\pt<4\GeV$.  However, the data may indicate a slight difference in suppression levels at higher \pt.  As the centrality range examined becomes more peripheral, the oscillating shape of \raaStar becomes less pronounced.  In the most peripheral collisions examined, the \XeXe data are relatively flat, indicating that the spectral shape for peripheral centrality ranges is similar to that of $\Pp\Pp$ collisions.  Although there is a large normalization uncertainty, the \raaStar is significantly below unity in this centrality range.  Such a suppression in peripheral events is not expected to be caused by strong energy loss effects, but might be related to correlations between the charged-particle yields in the mid-rapidity region with event activity in the range $3<\abs{\eta}<5.2$ that is used to determine the event centrality~\cite{Morsch:2017brb}.  Recent measurements of \raa in peripheral PbPb collisions by the ALICE Collaboration show a similar effect that has been interpreted as a bias caused by event selection and collision geometry~\cite{Acharya:2018njl}.  Studies in MB \textsc{hydjet} indicate this bias could be as large as 50\% at high \pt in the 70--80\% centrality range, but is expected to be less than 10\% for more central events.  This peripheral suppression could also be caused by a bias in \TAA values if the spatial distribution of hard partons inside each nucleus is narrower than expected~\cite{Jia:2009mq}.

The difference in the suppression between \raa for \PbPb collisions and \raaStar in \XeXe collisions can be directly compared with the ratio \rxepb.  Using the \PbPb charged-particle spectra from Ref.~\cite{Khachatryan:2016odn}, this quantity is determined for five centrality ranges and shown in Fig.~\ref{fig:XeXevsPbPb}.  The dark red box around unity shows the relative normalization uncertainty in the results.  The MC-based $\Pp\Pp$ extrapolation factor used in the construction of \raaStar is represented by the blue line, and shows the expected deviation of \rxepb from unity resulting from the different center-of-mass energies of the two collision systems.  In central events, the data for charged particles having $\pt<4\GeV$ are consistent with this expectation.  However, there is a sudden rise in \rxepb in the range of $5<\pt<10\GeV$, up to a value of 1.45.  This excess does not appear to be caused by the center-of-mass energy dependence and is located in the \pt region where \raaStar is the most suppressed.  This suggests a difference in the strength of energy loss in the two collision systems, which could be caused by the difference in the system size.  As the \pt increases towards $100\GeV$, the data slowly converge towards the values expected from the difference in the center-of-mass energy.  As the centrality range examined becomes more peripheral, the excess seen around 5 to 10\GeV decreases in strength.  In the most peripheral bins, \rxepb is consistent with the difference expected because of the center-of-mass energies throughout the entire \pt range.

\begin{figure}
  \centering
    \includegraphics[width=0.6\textwidth]{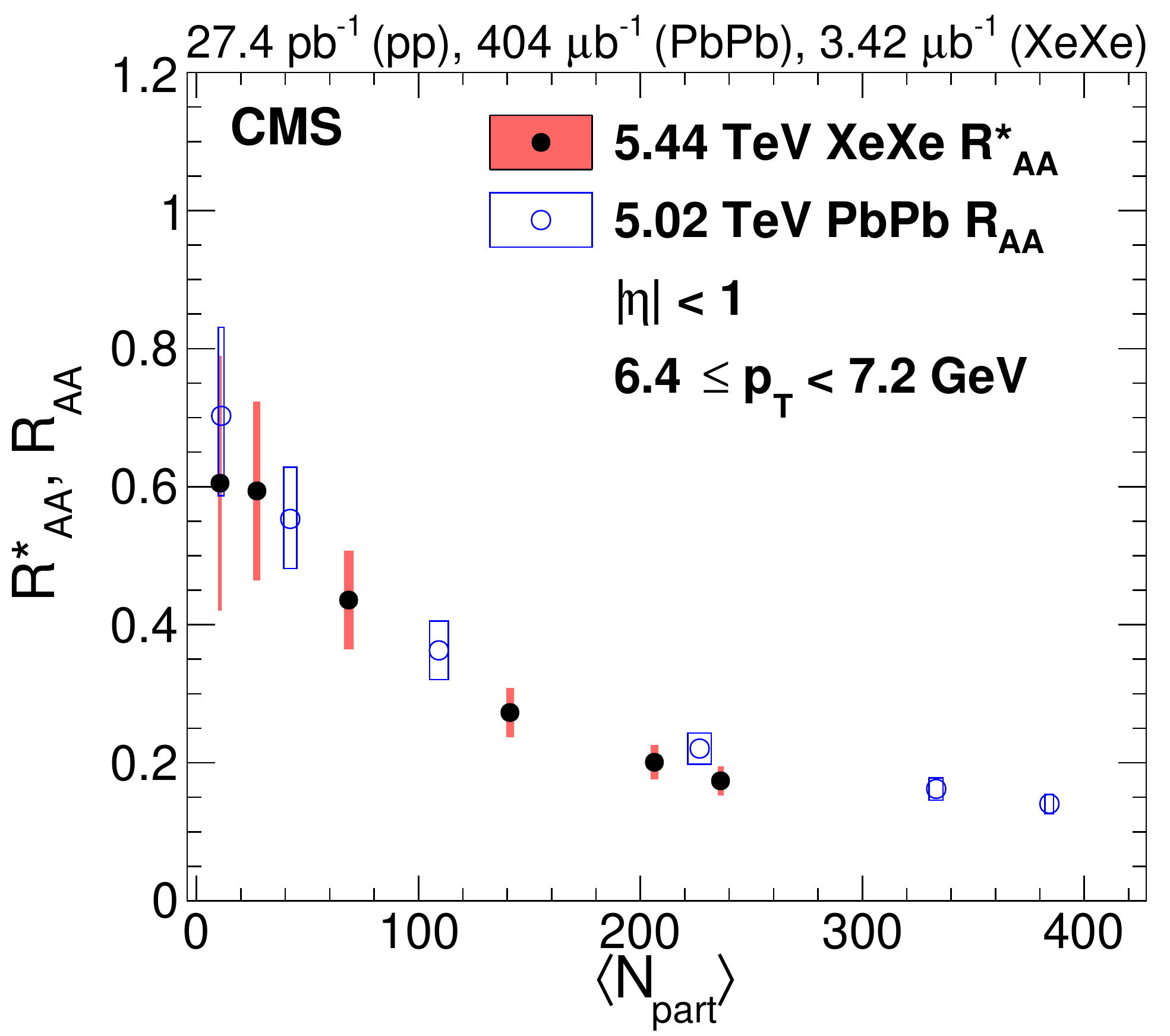}
  \caption{The charged-particle \raaStar for \XeXe collisions at $\sqrtsNN=5.44\TeV$ and \raa for \PbPb collisions at $5.02\TeV$, as a function of \avgNPart.  The solid pink and open blue boxes represent the total systematic uncertainties in the \XeXe and \PbPb data, respectively.}
  \label{fig:NpartSummary}
\end{figure}

Because xenon ions are smaller than lead ions, collisions at the same centrality will contain a different numbers of participating nucleons.  To compare \XeXe and \PbPb collisions having a similar number of colliding nucleons, the values of \raa and \raaStar for $6.4\leq \pt < 7.2\GeV$ are shown as a function of \avgNPart in Fig.~\ref{fig:NpartSummary}.  The chosen \pt range corresponds to the minima of \raa and \raaStar.  The boxes surrounding the data points show the total systematic uncertainties in the measurements.  The \raa and \raaStar values seem to follow a similar trend versus \avgNPart.  In particular, the values of \raa and \raaStar around $\langle N_{\text{part}} \rangle=220$ are compatible within the uncertainties.

\begin{figure}
  \centering
    \includegraphics[width=0.49\textwidth]{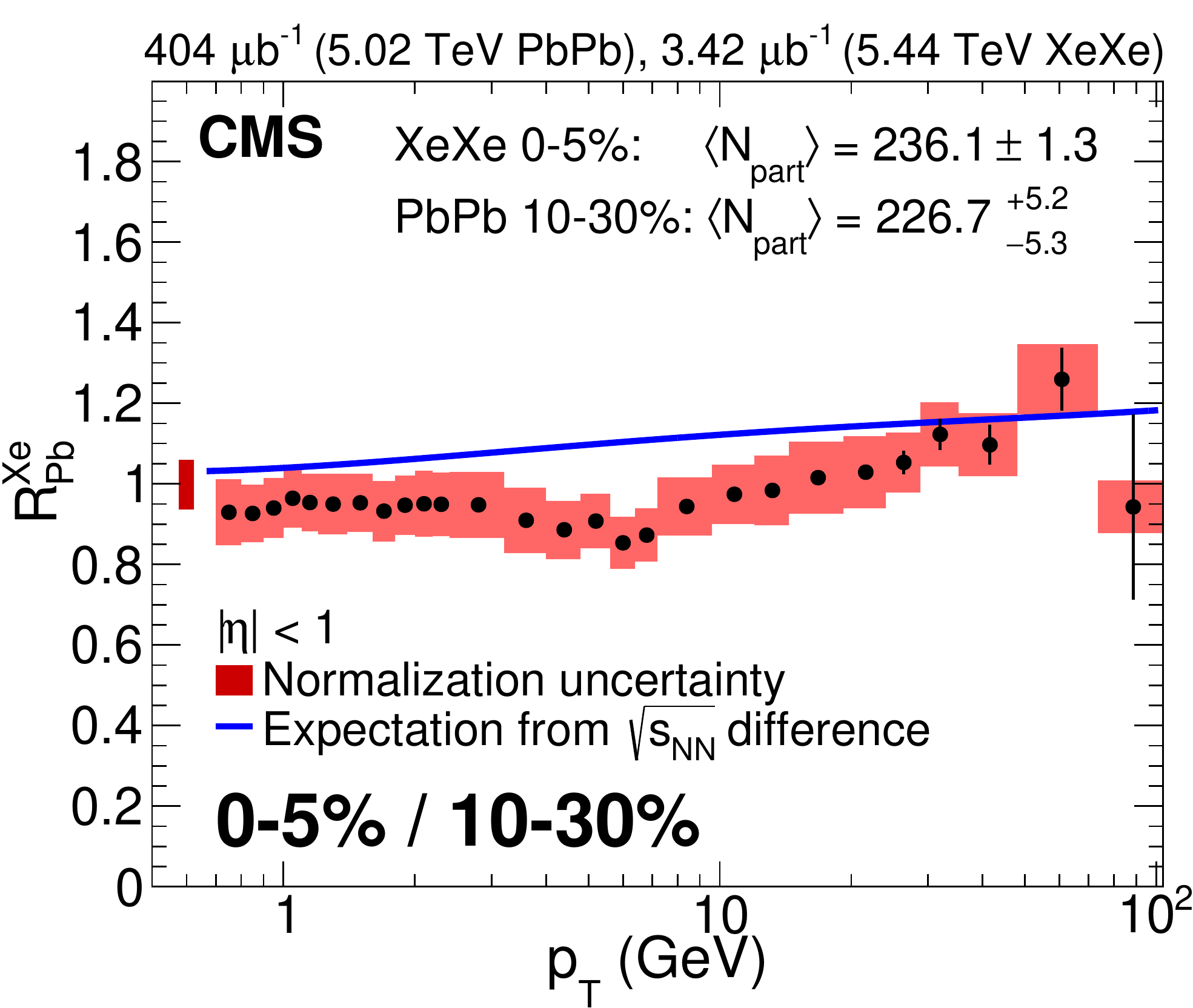}
    \includegraphics[width=0.49\textwidth]{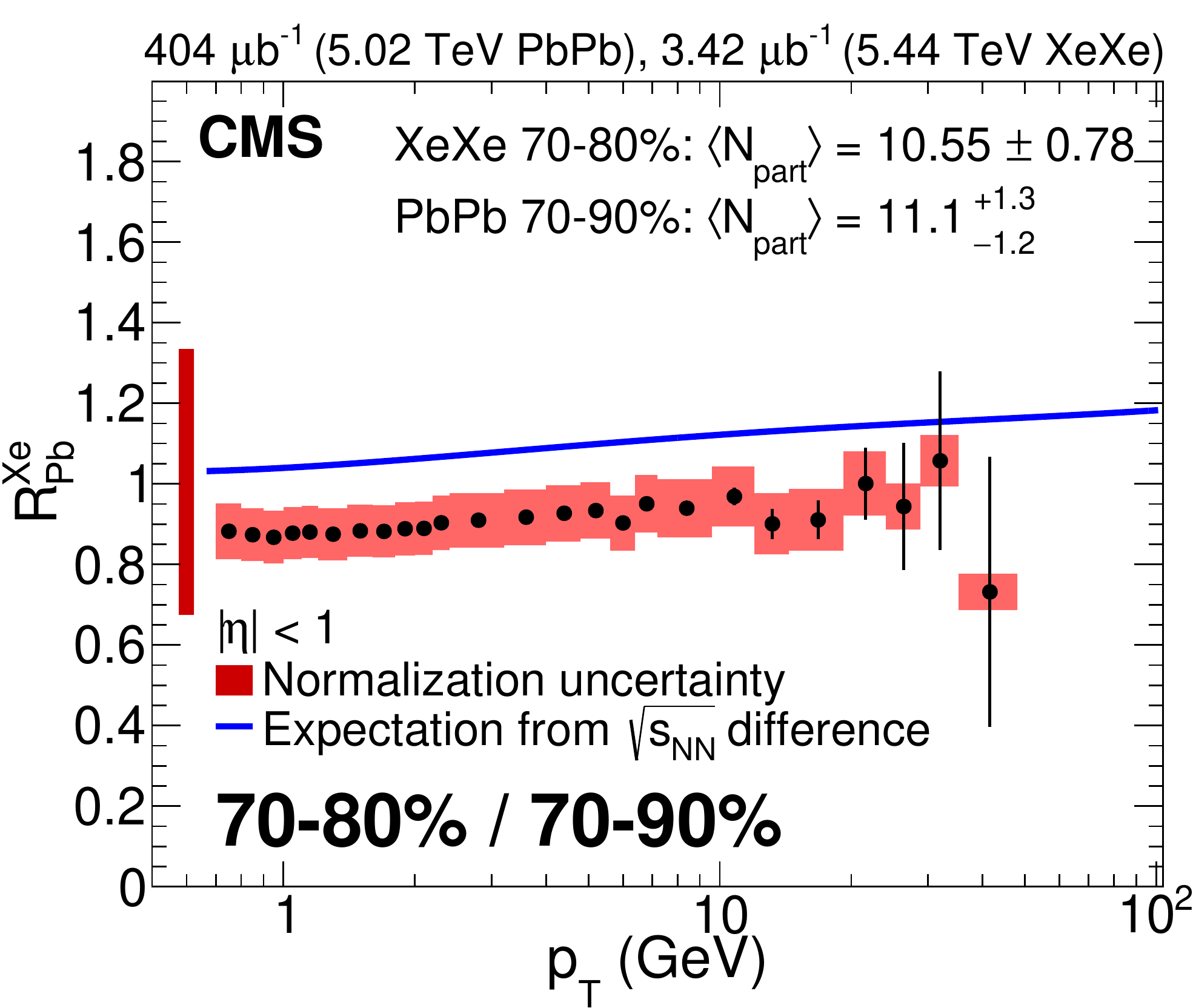}
  \caption{Measurements of \rxepb comparing centrality ranges having similar values of \avgNPart.  The blue line represents the expected deviation from unity caused by the different center-of-mass energies of the two collision systems.  The solid pink boxes represent the systematic uncertainties.}
  \label{fig:XeXevsPbPb_sameNPart}
\end{figure}

Measurements of \rxepb that compare data having similar \avgNPart, rather than centrality, are shown in Fig.~\ref{fig:XeXevsPbPb_sameNPart}.  The left panel compares 0--5\% \XeXe collisions with 10--30\% \PbPb collisions, which have \avgNPart values of $236.1\pm1.3$ and $226.7^{+5.2}_{-5.3}$, respectively.  In this case, the \rxepb values are slightly below the expectation from the different center-of-mass energies for $\pt < 20\GeV$, but are compatible with the expectation at higher \pt.  In the \pt range of 3--8\GeV, this ratio exhibits a slightly decreasing trend instead of the sharp rise seen when comparing similar centrality bins, reinforcing the conclusion that such a rise is due to a difference in the system size.  The right plot compares 70--80\% \XeXe events with 70--90\% \PbPb events.  In these centrality ranges, the \avgNPart value is $10.55\pm0.78$ for \XeXe collisions and $11.1^{+1.3}_{-1.2}$ for \PbPb events. The measurement has a large normalization uncertainty, but the shape of the distribution is very similar to the trend given by the center-of-mass energy difference of the two systems.

\begin{figure}
\centering
    \includegraphics[width=0.49\textwidth]{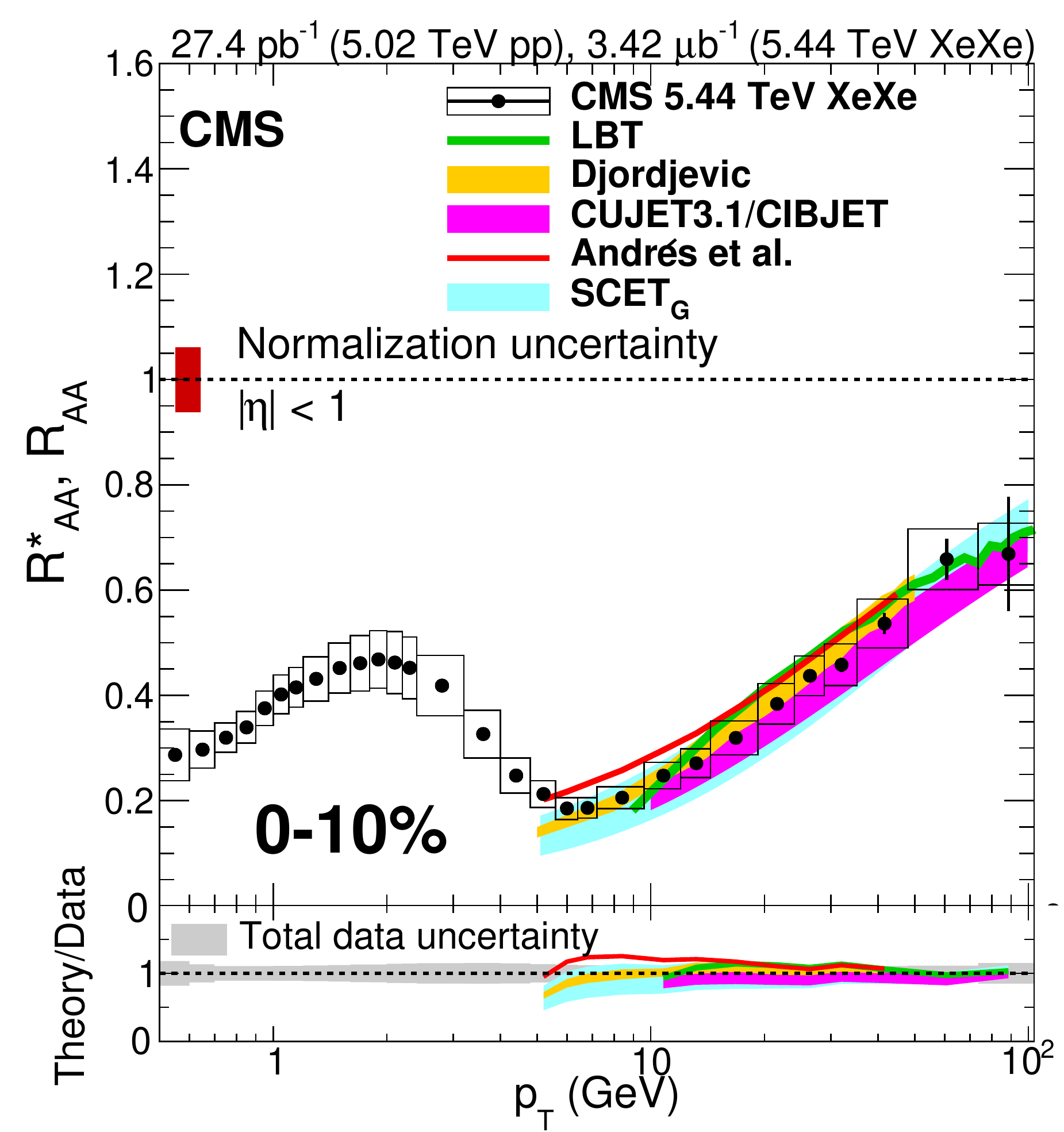}
    \includegraphics[width=0.49\textwidth]{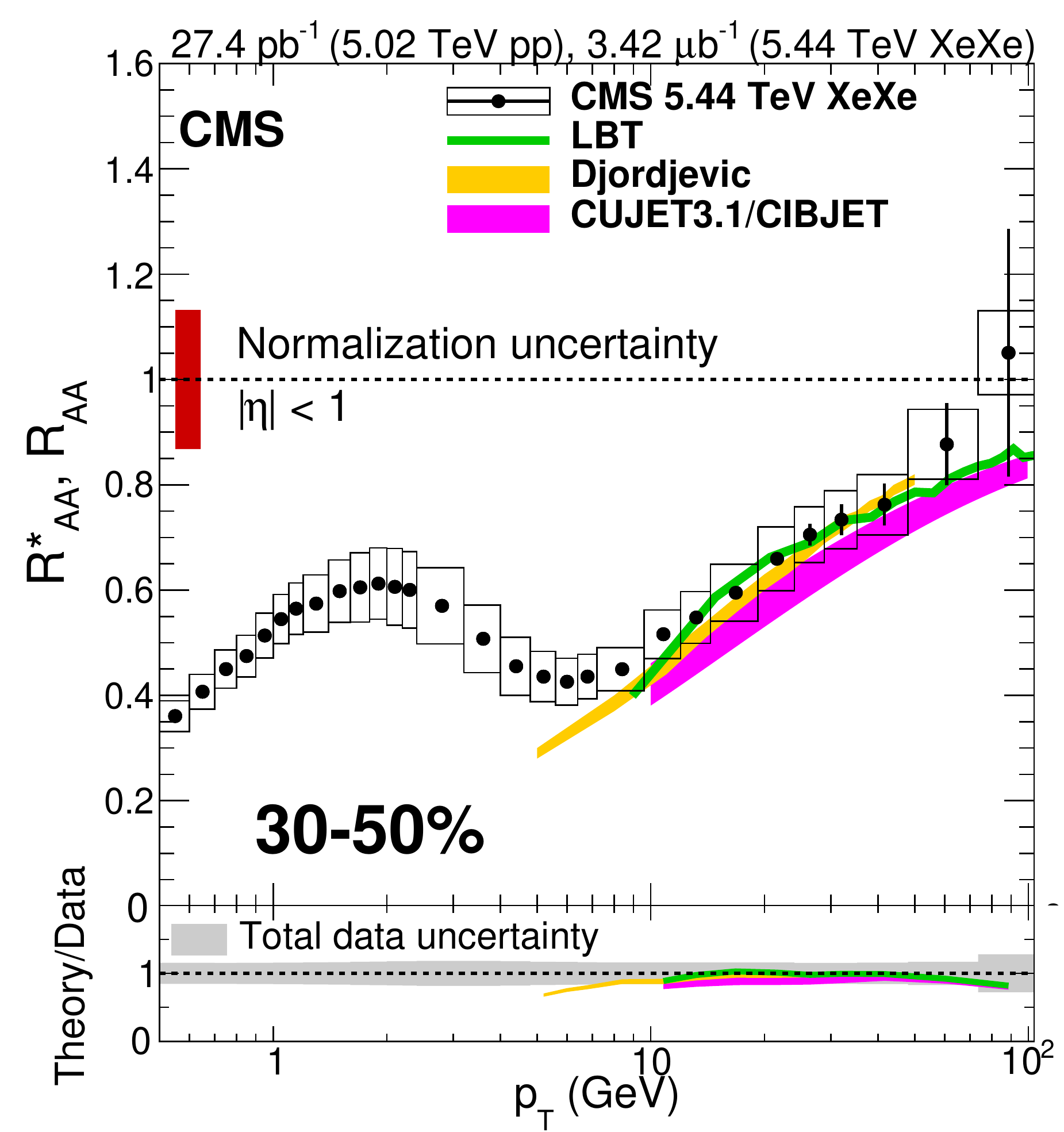}
  \caption{A comparison of the charged-particle \raaStar for \XeXe collisions at $\sqrtsNN=5.44\TeV$ with theoretical predictions from Refs.~\cite{He:2015pra,Cao:2017hhk,Kang:2014xsa,Chien:2015vja,Shi:2018lsf,Xu:2015bbz,Andres:2016iys,Djordjevic:2009cr,Djordjevic:2013xoa} for 0--10\% (left) and 30--50\% (right) centrality classes. The hollow black boxes represent the systematic uncertainties of the \XeXe data.  Ratios are shown in the bottom panels, where the gray band represents the total uncertainty in the measurement.}
  \label{fig:Theory_0_10}
\end{figure}

The \raaStar values in the 0--10\% and 30--50\% ranges are compared to various theoretical models in Fig.~\ref{fig:Theory_0_10}.  A ratio of each model to the data is provided in the bottom panels of the figure.  The green lines show the predictions of a linear Boltzmann transport (LBT) model of jet quenching, which uses the CLV$_{\textsc{isc}}$ hydrodynamics model for medium evolution~\cite{He:2015pra,Cao:2017hhk}.  This model predicts a quadratic path-length dependence of energy loss in a static medium.  It lies on the upper edge of the systematic uncertainty of the 0--10\% measurement in the range $20 < \pt < 60 \GeV$, but otherwise agrees with the data well.  The orange band is a model by Djordjevic that uses a dynamical energy loss formalism~\cite{Djordjevic:2009cr,Djordjevic:2013xoa}.   For this model, the medium undergoes Bjorken expansion, and the path-length dependence of energy loss is expected to be between linear and quadratic.  The prediction is compatible with the data in both centrality ranges except around $\pt=5\GeV$, where it is slightly below the data. The magenta region represents the prediction from \textsc{cujet3.1/cibjet} model, which incorporates two components~\cite{Xu:2015bbz,Shi:2018lsf}.  The first is a jet quenching model (\textsc{cujet3.1}) that includes the suppression of quark and gluon degrees of freedom and the emergence of chromo-magnetic monopole degrees of freedom. The second component, the \textsc{cibjet} framework, calculates the dependence of correlations between soft and hard azimuthal flow harmonics on an event-by-event basis. This model describes the 0--10\% data well, but lies on the lower edge of the data's uncertainty in the 30--50\% centrality range.  The red line shows a prediction of Andr\'{e}s~\etal that uses a `quenching weights' formalism to estimate the behavior of the medium transport coefficient, $\hat{q}$~\cite{Andres:2016iys}.  The evolution of the medium in this model is done with EKRT event-by-event hydrodynamics.  The prediction tends to agree with the top edge of the data's uncertainty range in the 0--10\% centrality range.  The light blue band shows a prediction from soft-collinear effective theory with Glauber gluons (SCET$_\textsc{G}$)~\cite{Kang:2014xsa,Chien:2015vja}.  The evolution of the background medium is modeled with the \textsc{iebe} hydrodynamics package.  In this model, \raa is found to scale roughly as $N_{\text{part}}^{2/3}$.  For central events, it slightly underestimates the \raaStar around 5\GeV, but generally agrees with the data.

\section{Summary}
\label{sec:summary}
The transverse momentum, \pt, spectra of charged particles in the pseudorapidity range $\abs{\eta}<1$ are measured in several ranges of collision centrality for xenon-xenon (\XeXe) collisions at a center-of-mass energy per nucleon pair of 5.44\TeV.  A proton-proton ($\Pp\Pp$) reference spectrum for the same energy is extrapolated from an existing measurement at $\sqrts=5.02\TeV$ using a scaling function calculated from simulated \PYTHIA events.  The nuclear modification factor with extrapolated reference, \raaStar, is constructed from these spectra.  In central events, \raaStar has a value of 0.17 in the \pt range of 6--8\GeV, before increasing to a value of around 0.7 at 100\GeV.  This suppression is less than what has been observed in a matching centrality range of lead-lead (\PbPb) collisions at a center-of-mass energy per nucleon pair of 5.02\TeV, even when accounting for the difference in collision energy.  In contrast, charged-particle production in \XeXe collisions is found to be slightly more suppressed than in \PbPb collisions that have a similar number of participating nucleons rather than a similar centrality.  Taken together, these observations illustrate the importance that collision system size and geometry have on the strength of parton energy loss.  Predictions from the Djordjevic, SCET$_\textsc{G}$ and \textsc{cujet3.1/cibjet} models are found to agree with the measured \raaStar.  The model of Andr\'{e}s \etal lies on the upper edge of the systematic uncertainties of \raaStar for central events.  Finally, calculations using a linear Boltzmann transport model also agree with the data, except for the kinematic range $15 < \pt < 40 \GeV$ in central events, where they follow the upper edge of the data's uncertainty.  These measurements help elucidate the nature of parton energy loss in \XeXe collisions and constrain the system size dependence of hot nuclear medium effects.

\begin{acknowledgments}
We congratulate our colleagues in the CERN accelerator departments for the excellent performance of the LHC and thank the technical and administrative staffs at CERN and at other CMS institutes for their contributions to the success of the CMS effort. In addition, we gratefully acknowledge the computing centres and personnel of the Worldwide LHC Computing Grid for delivering so effectively the computing infrastructure essential to our analyses. Finally, we acknowledge the enduring support for the construction and operation of the LHC and the CMS detector provided by the following funding agencies: BMBWF and FWF (Austria); FNRS and FWO (Belgium); CNPq, CAPES, FAPERJ, FAPERGS, and FAPESP (Brazil); MES (Bulgaria); CERN; CAS, MoST, and NSFC (China); COLCIENCIAS (Colombia); MSES and CSF (Croatia); RPF (Cyprus); SENESCYT (Ecuador); MoER, ERC IUT, and ERDF (Estonia); Academy of Finland, MEC, and HIP (Finland); CEA and CNRS/IN2P3 (France); BMBF, DFG, and HGF (Germany); GSRT (Greece); NKFIA (Hungary); DAE and DST (India); IPM (Iran); SFI (Ireland); INFN (Italy); MSIP and NRF (Republic of Korea); MES (Latvia); LAS (Lithuania); MOE and UM (Malaysia); BUAP, CINVESTAV, CONACYT, LNS, SEP, and UASLP-FAI (Mexico); MOS (Montenegro); MBIE (New Zealand); PAEC (Pakistan); MSHE and NSC (Poland); FCT (Portugal); JINR (Dubna); MON, RosAtom, RAS, RFBR, and NRC KI (Russia); MESTD (Serbia); SEIDI, CPAN, PCTI, and FEDER (Spain); MOSTR (Sri Lanka); Swiss Funding Agencies (Switzerland); MST (Taipei); ThEPCenter, IPST, STAR, and NSTDA (Thailand); TUBITAK and TAEK (Turkey); NASU and SFFR (Ukraine); STFC (United Kingdom); DOE and NSF (USA).

\hyphenation{Rachada-pisek} Individuals have received support from the Marie-Curie program and the European Research Council and Horizon 2020 Grant, contract No. 675440 (European Union); the Leventis Foundation; the A. P. Sloan Foundation; the Alexander von Humboldt Foundation; the Belgian Federal Science Policy Office; the Fonds pour la Formation \`a la Recherche dans l'Industrie et dans l'Agriculture (FRIA-Belgium); the Agentschap voor Innovatie door Wetenschap en Technologie (IWT-Belgium); the F.R.S.-FNRS and FWO (Belgium) under the ``Excellence of Science - EOS" - be.h project n. 30820817; the Ministry of Education, Youth and Sports (MEYS) of the Czech Republic; the Lend\"ulet (``Momentum") Programme and the J\'anos Bolyai Research Scholarship of the Hungarian Academy of Sciences, the New National Excellence Program \'UNKP, the NKFIA research grants 123842, 123959, 124845, 124850 and 125105 (Hungary); the Council of Science and Industrial Research, India; the HOMING PLUS programme of the Foundation for Polish Science, cofinanced from European Union, Regional Development Fund, the Mobility Plus programme of the Ministry of Science and Higher Education, the National Science Center (Poland), contracts Harmonia 2014/14/M/ST2/00428, Opus 2014/13/B/ST2/02543, 2014/15/B/ST2/03998, and 2015/19/B/ST2/02861, Sonata-bis 2012/07/E/ST2/01406; the National Priorities Research Program by Qatar National Research Fund; the Programa Estatal de Fomento de la Investigaci{\'o}n Cient{\'i}fica y T{\'e}cnica de Excelencia Mar\'{\i}a de Maeztu, grant MDM-2015-0509 and the Programa Severo Ochoa del Principado de Asturias; the Thalis and Aristeia programmes cofinanced by EU-ESF and the Greek NSRF; the Rachadapisek Sompot Fund for Postdoctoral Fellowship, Chulalongkorn University and the Chulalongkorn Academic into Its 2nd Century Project Advancement Project (Thailand); the Welch Foundation, contract C-1845; and the Weston Havens Foundation (USA).
\end{acknowledgments}
\bibliography{auto_generated}

\cleardoublepage \appendix\section{The CMS Collaboration \label{app:collab}}\begin{sloppypar}\hyphenpenalty=5000\widowpenalty=500\clubpenalty=5000\vskip\cmsinstskip
\textbf{Yerevan Physics Institute, Yerevan, Armenia}\\*[0pt]
A.M.~Sirunyan, A.~Tumasyan
\vskip\cmsinstskip
\textbf{Institut f\"{u}r Hochenergiephysik, Wien, Austria}\\*[0pt]
W.~Adam, F.~Ambrogi, E.~Asilar, T.~Bergauer, J.~Brandstetter, M.~Dragicevic, J.~Er\"{o}, A.~Escalante~Del~Valle, M.~Flechl, R.~Fr\"{u}hwirth\cmsAuthorMark{1}, V.M.~Ghete, J.~Hrubec, M.~Jeitler\cmsAuthorMark{1}, N.~Krammer, I.~Kr\"{a}tschmer, D.~Liko, T.~Madlener, I.~Mikulec, N.~Rad, H.~Rohringer, J.~Schieck\cmsAuthorMark{1}, R.~Sch\"{o}fbeck, M.~Spanring, D.~Spitzbart, A.~Taurok, W.~Waltenberger, J.~Wittmann, C.-E.~Wulz\cmsAuthorMark{1}, M.~Zarucki
% ...
\vskip\cmsinstskip
% ...
1:  Also at Vienna University of Technology, Vienna, Austria\\
% ...
\end{sloppypar}
\end{document}